\documentclass[a4paper,UKenglish]{lipics-v2016}

\usepackage[utf8]{inputenc}

\usepackage{alltt}
\usepackage{hyperref}
\usepackage{latexsym}
\usepackage{amssymb}
\usepackage{amsmath}
\usepackage{amsthm}
\usepackage[dvipsnames]{xcolor}

\usepackage{pgf}
\usepackage{tikz}
\usetikzlibrary{arrows,automata,shapes}

\newcommand{\sqsubsetneq}{\sqsubset}

\usepackage{alltt}

\newcommand{\ignore}[1]{}

\newcommand{\NN}{\mathbb{N}}

\definecolor{mycolor}{rgb}{0.99,0.78,0.07}

\newcommand{\PP}{\mathbb{P}}
\newcommand{\TT}{\mathbb{T}}
\newcommand{\QQ}{\mathbb{Q}}
\newcommand{\RR}{\mathbb{R}}
\renewcommand{\AA}{\mathcal{A}}

\newcommand{\bi}{\begin{itemize}}
\newcommand{\ei}{\end{itemize}}

\newcommand{\ind}{~\mathbb{I}~}

\newcommand{\pref}{\sqsubseteq}

\newcommand{\be}{\begin{equation}}
\newcommand{\ee}{\end{equation}}
\DeclareMathOperator{\dom}{dom}
\DeclareMathOperator{\view}{\partial}

\DeclareMathOperator{\dur}{dur}
\DeclareMathOperator{\last}{last}

\DeclareMathOperator{\plays}{plays}
\DeclareMathOperator{\alphabet}{Alph}
\DeclareMathOperator{\state}{state}
\DeclareMathOperator{\lcp}{lcp}
\DeclareMathOperator{\sstate}{strate}
\DeclareMathOperator{\decomp}{decomp}

\begin{document}

\bibliographystyle{plainurl}

\title{On the Control of Asynchronous Automata}

\iftrue
\titlerunning{On the Control of Asynchronous Automata} 

\author[1]{Hugo Gimbert}
\affil[1]{LaBRI, CNRS, Universit\'e de Bordeaux, France\\ \tt{hugo.gimbert@cnrs.fr}}

\authorrunning{H. Gimbert}

\Copyright{Hugo Gimbert}

\subjclass{B.1.2 Automatic synthesis, H.3.4 Distributed systems}
\keywords{
asynchronous automata, 
Controller synthesis
}

\EventEditors{John Q. Open and Joan R. Acces}
\EventNoEds{2}
\EventLongTitle{42nd Conference on Very Important Topics (CVIT 2016)}
\EventShortTitle{CVIT 2016}
\EventAcronym{CVIT}
\EventYear{2016}
\EventDate{December 24--27, 2016}
\EventLocation{Little Whinging, United Kingdom}
\EventLogo{}
\SeriesVolume{42}
\ArticleNo{23}

\fi

\maketitle

\begin{abstract}
The decidability of the distributed version of the Ramadge and Wonham controller synthesis problem~\cite{ramadge1989control},
where both the plant and the controllers are modeled as asynchronous automata~\cite{zautomata,thebook}
and the controllers have causal memory
is a challenging open problem~\cite{alook,mumu}.
There exist three classes of plants for which the existence of a correct controller with causal memory has been shown decidable: when the dependency graph of actions is series-parallel, 
when the processes are connectedly communicating and when the dependency graph of processes is a tree. 
We design a class of plants, called decomposable games, 
with a decidable controller synthesis problem.
This provides
 a unified proof of the three existing decidability results
 as well as new examples of decidable plants.
 \end{abstract}

\section{Introduction}

The decidability of the distributed version of the Ramadge and Wonham control problem~\cite{ramadge1989control},
where both the plant and the controllers are modeled as asynchronous automata~\cite{zautomata,thebook}
and the controllers have causal memory
is a challenging open problem.
Very good introductions to this problem are given in~\cite{alook,mumu}.

In this setting a controllable plant is distributed on several finite-state processes
which interact asynchronously using shared actions.
On every process, the local controller can choose to block some of the actions,
called \emph{controllable} actions, but it cannot block the \emph{uncontrollable} actions from the environment.
The choices of the local controllers are based on two sources of information.
\begin{itemize}
\item
First the controller monitors the sequence of states and actions of the local process.
This information is called the \emph{local view} of the controller.
\item
Second when a shared action is played by several processes
then all the controllers of these processes can exchange as much information as they want.
In particular together they can compute their mutual view of the global execution:
their \emph{causal past}.
\end{itemize}

A  controller is correct if it guarantees that every
possible execution of the plant
satisfies some specification. The controller synthesis problem is a decision problem which, given a plant as input,
asks whether the system admits a correct controller.
In case such a controller exists, the algorithm should as well compute one.

The difficulty of controller synthesis depends on several factors, e.g.:
\begin{itemize}
\item the size and architecture (pipeline, ring, ...) of the system,
\item the information available to the controllers,
\item the specification.
\end{itemize}
Assuming that processes can exchange information upon synchronization and use their causal past to take decisions is one of the key aspects to get decidable synthesis problems~\cite{gastin}.
In early work on distributed controller synthesis,
for example in the setting of~\cite{pneuli1990distributed}, the only source of information available to the controllers is their local view.
In this setting, distributed synthesis is not decidable in general, except for very particular architectures like the pipeline architecture. The paper~\cite{finkbeiner2005uniform} proposes information forks as an uniform notion explaining the (un)decidability results in distributed synthesis. The idea of using causal past as a second source of information appeared in~\cite{gastin}.

\medskip

We adopt a modern terminology and call the plant a \emph{distributed game} and the controllers are \emph{distributed strategies} in this game.
A distributed strategy is a function that maps the causal past of processes
to a subset of controllable actions.
In the present paper we focus on the \emph{termination condition},
which is satisfied when each process is guaranteed to terminate its computation in finite time,  in a final state.
A distributed strategy is winning if it guarantees the termination condition,
whatever uncontrollable actions are chosen by the environment.

We are interested in the following problem, whose decidability is an open question.
\smallskip
\newcommand{\dsp}{{\sc distributed synthesis problem}}

\noindent \dsp: given a distributed game decide
whether  there exists a winning strategy.

\smallskip

There exists three classes of plants for which the \dsp\ 
has been shown decidable:
\begin{enumerate}
\item
when the dependency graph of actions is series-parallel~\cite{gastin}, 
\item
when the processes are connectedly communicating~\cite{madhu},
\item
 and when the dependency graph of processes is a tree~\cite{acyclic,DBLP:conf/fsttcs/MuschollW14}. 
\end{enumerate}

A series-parallel game is a game such that
the dependency graph $(A,D)$ of the alphabet $A$
is a co-graph.
Series-parallel games were proved decidable in~\cite{gastin}, for a different setup than ours:
in the present paper we focus on process-based control while~\cite{gastin} was focusing
on action-based control. Actually action-based control is more general than process-based
control, see~\cite{alook} for more details. The results of the present paper could probably be extended to action-based control however we prefer to stick to process-based control in order to keep the model intuitive.
To our knowledge, the result of~\cite{gastin} was the first discovery of a class of asynchronous distributed system with causal memory for which the \dsp\ is decidable

Connectedly communicating games have been introduced~\cite{madhu}.
A game is connectedly communicating if there is a bound $k$ such that if a process $p$ executes $k$ steps in parallel of another process $q$ then all further actions of $p$ will be parallel to $q$.
The event structure of a 
connectedly communicating games has a decidable MSO theory~\cite{madhu}
which implies that the \dsp\ is decidable for these games.

An acyclic game is a game where processes are arranged as a tree and actions are either local or synchronize a father and its son.
Even in this simple setting the \dsp\ is non-elementary hard~\cite{acyclic}.

\paragraph*{Our contribution}
We develop a new proof technique to address the \dsp,
and provide a unified proof of decidability for series-paralell,
connectedly communicating and acyclic games.
We design a class of games, called \emph{decomposable games}, for which the \dsp\ is decidable.
This leads to new examples of decidable architectures for controller synthesis.

The winning  condition of the present paper is the termination of all processes in a final state.
Richer specifications can be expressed by parity conditions.
In the present paper we stick to termination conditions for two reasons.
First,
the long-term goal of this research is to establish the decidability or undecidability
of the distributed controller synthesis problem. A possible first step is to prove decidability
for games with termination conditions.
Second,
it seems that the results of the present paper can be lifted to parity games,
using the same concepts but at the cost of
some extra technical details
needed to reason about infinite plays.

Our proof technique consists in simplifying a winning strategy by looking for useless parts
to be removed in order to get a smaller winning strategy. These parts are called \emph{useless repetitions}. Whenever a useless repetition exists, we remove it using an operation called a \emph{shortcut} in order to get a simpler strategy. Intuitively, a shortcut is  a kind of cut-and-paste operation
which makes the strategy smaller. By taking shortcuts again and again, we make the strategy smaller and smaller, until it does not have any useless repetition anymore.

If a winning strategy exists, there exists one with no useless repetition. In decomposable games, there is a computable upper bound on the size of strategies with no useless repetition, which leads to decidability of the controller synthesis problem.

Performing cut-and-paste in a distributed game is not as easy as doing it in a single-process game.
In a single-process game,
strategies are trees and one can cut a subtree from a node A and paste it to any other node B, and the operation makes sense as long as the state of the process in the same in both  A and B.
In the case of a general distributed strategy,
designing  cut-and-paste operations is more challenging. Such operations on the strategy tree  should be consistent with the level of information of each process, in order to preserve the fundamental property of distributed strategies: the decisions taken by a process should depend only of its causal view,
not on parallel events.

The decidability of series-parallel games established in~\cite{gastin} relies also on some simplification of the winning strategies,
in order to get \emph{uniform} strategies. 
The series-parallel assumption is used to guarantee that the result of the replacement of a part of a strategy by a uniform strategy  is still a strategy, as long as the states of all processes coincide. Here we work without the series-parallel assumption,
and matching the states is not sufficient for a cut-and-paste operation to be correct.
 
This is the reason for introducing the notion of \emph{lock}. 
A lock is a part of a strategy where an information is guaranteed to spread in a team of processes before any of these processes synchronize with a process outside the team.
When two locks A and B are similar, in some sense made precise in the paper, the lock B can be cut and paste on lock A. Upon arrival on A, a process of the team initiates a change of strategy, which progressively spreads across the team. All processes of the team should eventually 
play as if the 
play
 from A to B had  already taken place, although it actually did not.


The complexity of our algorithm is really bad, so probably this work has no immediate practical applications. This is not surprising
since the problem is non-elementary even for the class of acyclic games~\cite{acyclic}.
Nevertheless we think this paper
sheds new light on the difficult open problem of distributed synthesis.

\paragraph*{Organization of the paper}

Section~2 introduces the \dsp.
Section~\ref{sec:examples} provides several examples. In section~\ref{sec:simplifying} we show how to simplify strategies which contains useless repetitions,
and prove that if a winning strategy exists, there exists one without any useless repetition. Finally, section~\ref{sec:decomposable} introduces the class of decomposable games and show their controller synthesis problem is decidable.
Missing proofs can be found in the appendix. 

\section{The distributed synthesis problem}

The theory of Mazurkiewicz traces is very rich,
for a thorough presentation see~\cite{thebook}.
Here we only fix notations and recall the notions of traces, views, prime traces and parallel traces.

We fix an alphabet $A$ and a symmetric and reflexive dependency relation $D \subseteq A\times A$
and the corresponding independency relation $\ind \subseteq A\times A$ defined as $
\forall a,b\in A, (a \ind b) \iff (a,b)\not\in D$.
A  \emph{Mazurkiewicz trace}
or, more simply, a \emph{trace},
is an equivalence class
for the smallest equivalence relation $\equiv$ on $A^*$ 
which commutes independent letters i.e.
for every letters $a,b$
and every words $x,y$,
\[
a \ind b \implies
xaby\equiv
xbay\enspace. 
\]
The words in the equivalence class are the \emph{linearizations} of the trace.
The trace whose only linearization is the empty word is 
denoted $\epsilon$.
All linearizations of a trace $u$ have the same set of letters and length, denoted respectively $\alphabet(u)$ and $|u|$.
Given $B\subseteq A$,
the set of traces such that
$\alphabet(u)\subseteq B$
is denoted $B_\equiv^*$
in particular the set of all traces is 
$A^*_\equiv$.

The concatenation on words naturally extends to traces.
Given two traces $u,v\in A^*_\equiv$, the trace $uv$ is the equivalence class of any word in $uv$.
The prefix relation $\pref$ is defined by:
\[
(u\pref v \iff \exists w \in A^*_\equiv, uw= v)\enspace.
\]

\paragraph*{Maxima, prime traces and parallel traces}
A letter $a\in A$ is a \emph{maximum} of a trace $u$ if it is the last letter of one of the linearizations of $u$ .
A trace $u\in A_\equiv^*$ is \emph{prime} if it has a unique maximum,
denoted $\last(u)$
and called the last letter of $u$.
Two prime traces $u$ and $v$ are said to be \emph{parallel}
if 
\begin{itemize}
\item
neither 
$u$ is a prefix of $v$ nor 
$v$ is a prefix of $u$; and
\item
there is a trace $w$ such that both $u$ and $v$ are prefixes of $w$.
\end{itemize}

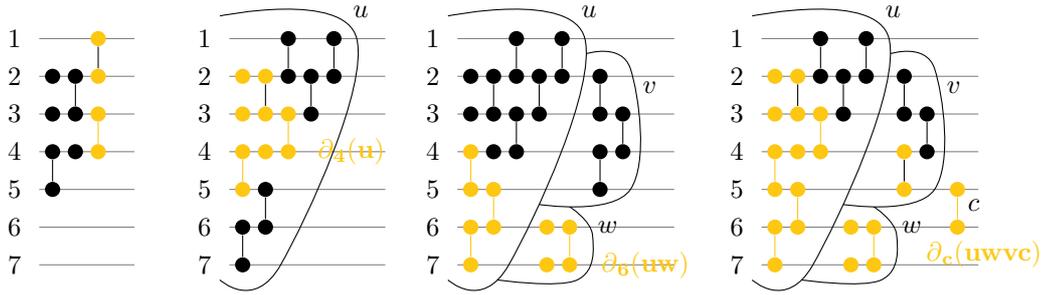
\begin{figure}

\begin{tikzpicture}
\begin{scope}[scale=1]

\newcommand{\eeee}{mycolor}
\newcommand{\eeeeee}{mycolor}
\newcommand{\eeeee}{1.5cm}
\newcommand{\vfour}{black}
\newcommand{\vfourr}{mycolor}
\newcommand{\colw}{black}
\newcommand{\spec}{\vfour}
\newcommand{\ff}[1]{{}}

\newcommand{\ssss}{
\tikzstyle{every node}=[node distance=.5cm]
\node(lab1) at (1,-1) {$1$ };
\node(lab2) [below of=lab1] {$2$ };
\node(lab3) [below of=lab2] {$3$ };
\node(lab4) [below of=lab3] {$4$ };
\node(lab5) [below of=lab4] {$5$ };
\node(lab6) [below of=lab5] {$6$ };
\node(lab7) [below of=lab6] {$7$ };
\tikzstyle{every node}=[node distance=.2cm]
\node(l1) [right of=lab1] {};
\tikzstyle{every node}=[node distance=.5cm]
\node(l2) [below of=l1] {};
\node(l3) [below of=l2] {};
\node(l4) [below of=l3] {};
\node(l5) [below of=l4] {};
\node(l6) [below of=l5] {};
\node(l7) [below of=l6] {};
\tikzstyle{every node}=[node distance=\eeeee]
\node(r1) [right of=l1] {};
\draw[gray] (l1) -- (r1);
\node(r2) [right of=l2] {};
\draw[gray] (l2) -- (r2);
\node(r3) [right of=l3] {};
\draw[gray] (l3) -- (r3);
\node(r4) [right of=l4] {};
\draw[gray] (l4) -- (r4);
\node(r5) [right of=l5] {};
\draw[gray] (l5) -- (r5);
\node(r6) [right of=l6] {};
\draw[gray] (l6) -- (r6);
\node(r7) [right of=l7] {};
\draw[gray] (l7) -- (r7);

\tikzstyle{every state}=[fill=black,draw=none,inner sep=0pt,minimum size=0.2cm]
\tikzstyle{every node}=[node distance=.3cm]
\node[state](a2)[right of=l2, fill=\vfour]{};
\tikzstyle{every node}=[node distance=.5cm]
\node[state](a3)[below of=a2, fill=\vfour]{};
\node[state](a4)[below of=a3, fill=\spec]{};
\node[state](a5)[below of=a4, fill=\spec]{};
\draw[\spec] (a4) -- (a5);
\ff{\node[state](a6)[below of=a5, fill=\colw]{};
\node[state](a7)[below of=a6, fill=\colw]{};
\draw[\colw] (a6) -- (a7);}

\node[state](b2)[right of=a2, node distance=.3cm, fill=\vfour]{};
\node[state](b3)[below of=b2, fill=\vfour]{};
\draw (b2) -- (b3);
\node[state](b4)[below of=b3, fill=\vfour]{};
\ff{\node[state](b5)[below of=b4, fill=\eeeeee]{};
\node[state](b6)[below of=b5, fill=\eeeeee]{};
\draw[color=\eeeeee] (b5) -- (b6);
}

\node[state](c2)[right of=b2, fill=\eeee, node distance=.3cm]{};
\node[state](c1)[above of=c2, fill=\eeee]{};
\draw (c1) -- (c2);
\node[state](c3)[below of=c2, fill=\vfourr]{};
\node[state](c4)[below of=c3, fill=\vfourr]{};
\draw[\vfourr] (c3) -- (c4);

\ff{
\node[state](d2)[right of=c2, node distance=.3cm]{};
\node[state](d3)[below of=d2]{};
\draw (d2) -- (d3);

\node[state](e2)[right of=d2, node distance=.3cm, fill=\eeee]{};
\node[state](e1)[above of=e2, fill=\eeee]{};
\draw[color=\eeee] (e2) -- (e1);

\node(u)[above right of=e1]{$u$};

\draw [black] plot [smooth, tension=0.7] coordinates { (1.2,-0.7) (3,-1) (2,-4) (1.2,-4.2)};
}
}
%
%
%

\ssss

\renewcommand{\eeee}{black}
\renewcommand{\eeeee}{2.3cm}
\renewcommand{\vfour}{mycolor}
\renewcommand{\vfourr}{\vfour}
\renewcommand{\eeeeee}{black}
\renewcommand{\ff}[1]{{#1}}

\begin{scope}[shift={(2.5,0)}]
\ssss
\node(view)[right of =c4, node distance=0.8cm]{\color{\vfour}$\bf\view_4(u)$};
\end{scope}

\renewcommand{\eeeee}{3.1cm}
\renewcommand{\vfour}{black}
\newcommand{\colv}{black}
\renewcommand{\colw}{mycolor}
\renewcommand{\spec}{\colw}
\renewcommand{\eeeeee}{\colw}
\newcommand{\eeeeeeee}{black}

\newcommand{\sssss}{
\ssss
\node[state](f2)[right of =e2,fill=\colv]{};
\node[state](f3)[below of =f2,fill=\colv]{};
\draw[\colv] (f2)--(f3);
\node[state](f4)[below of =f3,fill=\eeeeeeee]{};
\node[state](f5)[below of =f4,fill=\eeeeeeee]{};
\draw[\colv] (f4)--(f5);
\node[state](g3)[right of =f3, node distance=.3cm]{};
\node[state](g4)[below of =g3]{};
\draw[\colv] (g3)--(g4);

\node(v)[above right of =g3]{$v$};
\draw [black] plot [smooth, tension=0.7] coordinates { (3.02,-1.2) (3.6,-1.4) (3.6,-3) (2.4,-3.2)};

\node[state](d6)
[right of =b6,fill=\colw, node distance=.7cm]{};
\node[state](d7)
[below of =d6,fill=\colw]{};
\node[state](e6)
[right of =d6,fill=\colw, node distance=.3cm]{};
\node[state](e7)
[below of =e6,fill=\colw]{};
\draw[\colw] (e6)--(e7);
\node(w)
[right of =e6, node distance=.5cm]{$w$};

\draw [black] plot [smooth, tension=0.7] coordinates { (2.8,-3.23) (3.1,-3.6) (2.9,-4.2) (1.85,-4.2)};
}

\begin{scope}[shift={(5.5,0)}]
\sssss

\node(view)[right of =e7, node distance=1cm]{\color{\colw}$\bf\view_6(uw)$};

\end{scope}

\renewcommand{\eeeee}{3.1cm}
\renewcommand{\vfour}{mycolor}
\renewcommand{\colv}{black}
\renewcommand{\colw}{mycolor}
\renewcommand{\spec}{mycolor}
\renewcommand{\eeeeee}{mycolor}
\renewcommand{\eeeeeeee}{mycolor}

\begin{scope}[shift={(9.5,0)}]
\sssss

\node[state](oups)[right of =f5, node distance=.7cm, fill=mycolor]{};
\node(c)[below right of =oups, node distance=.3cm]{$c$};
\node[state](oups2)[below of =oups, fill=mycolor]{};
\draw[mycolor] (oups)--(oups2);

\node(w)
[below right of =oups2, node distance=.5cm]{ \color{mycolor}$\bf\view_c(uwvc)$};

\end{scope}

\end{scope}
\end{tikzpicture}

\caption{\label{fig:example}
The set processes is $\{1\ldots 7\}$.
A letter is identified with its domain.
Here the domains are either singletons, represented by a single dot,
or pairs of contigous processes,
represented by two dots connected with a vertical segment.
On the left handside is represented the trace
$
\{2\}\{3\}\{4,5\}\{2,3\}\{4\}\{1,2\}\{3,4\}
=
\{4,5\}\{4\}\{2\}\{3\}\{2,3\}\{3,4\}\{1,2\}
$
which has two maximal letters $\{1,2\}$ and $\{3,4\}$ thus is not prime.
Center left: process $4$ sees only its causal view $\view_4(u)$ (in yellow).
Center right: $uvw=uwv$ since $\dom(v)\cap \dom(w)=\emptyset$. Both $uv$ and $\view_{6}(uw)$ (in yellow) are prime prefixes of $uvw$ and they are parallel.
Right: $uv$ and $\view_{c}(uvwc)$ (in yellow) are parallel.
}
\end{figure}

These notions are illustrated on Fig.~\ref{fig:example}.

\paragraph*{Processes and automata}


Asynchronous automata are to traces what finite automata are to finite words, as witnessed by Zielonka's theorem~\cite{zautomata}. An asynchronous automaton is a collection of automata on finite words, whose transition tables do synchronize on certain actions. 


\begin{definition}
An asynchronous automaton on alphabet $A$ with processes $\PP$ is a tuple
$
\mathcal{A} = ((A_p)_{p\in \PP},(Q_p)_{p\in \PP},  (i_p)_{p\in \PP},(F_p)_{p\in \PP}, \Delta)
$
where:
\begin{itemize}
\item
 every process $p\in \PP$ has a set of actions
 $A_p$, a set of states $Q_p$
 and $i_p\in Q_p$ is the initial state of $p$ and $F_p\subseteq Q_p$ its set of final states.
\item
 $A=\bigcup_{p\in \PP} A_p$.
 For every letter $a\in A$, the domain of $a$ is 
 $
 \dom(a)=\{p\in \PP\mid a\in A_p\}\enspace.
$
 \item
$\Delta$ is a set of transitions of the form 
$(a,(q_p,q'_p)_{p\in\dom(a)})$
where $a\in A$
and $q_p,q'_p\in Q_p$.
Transitions are \emph{deterministic}: for every $a\in A$,
if
$\delta=(a,(q_p,q'_p)_{p\in \dom(a)})\in \Delta$
and $\delta'=(a,(q_p,q''_p)_{p\in \dom(a)})\in \Delta$
then $\delta=\delta'$
(hence 
 $\forall p \in \dom(a),q'_p=q''_p$).
\end{itemize}
\end{definition}

Such an automaton works asynchronously:
each time a letter $a$ is processed, the states of the processes
in $\dom(a)$ are updated according to the corresponding
transition, while the states of other processes do not change.
This induces a natural commutation relation $\ind$ on $A$:
two letters commute iff they have no process in common i.e.
\begin{align*}
& (a \ind b) \iff (\dom(a)\cap \dom(b) = \emptyset)\enspace.
\end{align*}

The set of \emph{plays} of the automaton $\AA$
is a set of traces denoted $\plays(\AA)$ and defined inductively, along with 
a mapping $\state:\plays(\AA) \to \Pi_{p\in\PP}Q_p$.
\begin{itemize}
\item $\epsilon$ is a play and $\state(\epsilon)=(i_p)_{p\in \PP}$,
\item for every play $u$ such that $(\state_p(u))_{p\in\PP}$ is defined
and $\left(a,(\state_p(u),q_p)_{p\in\dom(a)}\right)$
is a transition then
$ua$ is a play and
$
\forall p \in \PP,
\state_p(ua) =
\begin{cases}
\state_p(u) &\text{ if $p\not\in\dom(a)$,}\\
q_p &\text{ otherwise.}
\end{cases}
$
\end{itemize}

For every play $u$, $\state(u)$ is called the \emph{global state} of $u$.
The inductive definition of $\state(u)$ is correct because it is invariant by
commutation of independent letters of $u$.

\paragraph*{Counting actions of a process}
For every trace $u$ we can count how many times a process $p$
has played an action in $u$, which we denote $|u|_p$.
Formally, $|u|_p$ is first defined for words, as the length of the projection
of $u$ on $A_p$, which is invariant by commuting letters.
The domain of a trace is defined as 
\[
\dom(u) = \left\{ p \in \PP \mid |u|_p \neq 0\right\}\enspace.
\]

\paragraph*{Views, strategies and games}

Given an automaton $\AA$,
we want the processes to choose actions
which guarantee that
every play eventually terminates in a final state.

To take into account the fact that some actions are controllable by processes while some other actions are not, we assume that $A$ is partitioned in 
\[
A=A_c \sqcup A_e
\]
where $A_c$ is the set of controllable actions and
$A_e$ the set of (uncontrollable)  environment actions.
Intuitively, processes cannot prevent their environment to 
play actions in $A_e$, while they can decide whether to block or allow any action in $A_c$.

We adopt a modern terminology and call the automaton $\AA$ together with the partition
$A=A_c\sqcup A_e$ a \emph{distributed game}, or even more simply a \emph{game}.
In this game the processes play distributed strategies,
which are individual plans of action for each process.
The choice of actions by a process $p$ is dynamic: at every step, $p$ chooses a new set of controllable actions, depending on its  information about the way the play is going on. This information is limited since
processes cannot communicate together unless they synchronize on a
common action. In that case however they exchange as much information about the play as they want.
Finally, the information missing to a process is the set of actions which happened in parallel of its own actions.
The information which remains is called the $p$-view of the play,
and is defined formally as follows.

%

%

\begin{definition}[Views]\label{def:views}
For every set of processes $\QQ\subseteq \PP$
and trace $u$,
the \emph{$\QQ$-view} of $u$, denoted $\view_\QQ(u)$,
 is the unique trace such that
 $u$ factorizes as
$
u=\view_\QQ(u)\cdot v$ and
$v$ is the longest suffix of $u$ such that 
$\QQ\cap \dom(v)=\emptyset$.
%
%
In case $\QQ$ is a singleton $\{p\}$ the view is denoted
$\view_p(u)$ and is a prime trace.
For every letter $a \in A$ we denote $\view_a(u)=\view_{\dom(a)}(u)$.
%
%
\end{definition}

The well-definedness of the $\QQ$-view is shown in the appendix, where we also establish:
\begin{align}
\label{eq:viewdec}
&
\view_\QQ(uv)=\view_{\QQ'}(u)\view_\QQ(v)
\text{ where }
\QQ'=\QQ\cup \dom(\view_\QQ(v))\enspace\\
\label{viewsub}
& (\QQ\subseteq \QQ') \implies 
(\view_\QQ(u)
 \pref
\view_{\QQ'}(u))\enspace.
\end{align}

We can now define what is a distributed strategy.

\begin{definition}[Distributed strategies, consistent and maximal plays]
Let $G=(\AA,A_c, A_e)$ be a distributed game.
A \emph{strategy for process $p$} in $G$ is a mapping
which associates with every play $u$ a set of actions $\sigma_p(u)$ such that:
\begin{itemize}
\item
environment actions are allowed: $A_e\subseteq \sigma_p(u)$,
\item
the decision depends only on the view of the process:
$\sigma_p(u)=\sigma_p(\view_p(u))$.
\end{itemize}
A \emph{distributed strategy}
 is a tuple $\sigma=(\sigma_p)_{p\in\PP}$
where each $\sigma_p$ is a strategy of process $p$.
A play $u=a_1\cdots a_{|u|}\in\plays(\AA)$ is \emph{consistent with $\sigma$},
or equivalently is a \emph{$\sigma$-play} if:
\[
\forall i \in 1\ldots |u|,
\forall p\in\dom(a_i),
a_i\in\sigma_p(a_1\cdots a_{i-1})\enspace.
\]
A $\sigma$-play is \emph{maximal} if it is not the strict prefix of another $\sigma$-play.
\end{definition}

Note that a strategy is forced to allow every environment action to be executed at every moment.
This may seem to be a huge strategic advantage for the environment.
However depending on the current state,
not every action can be effectively used in a transition
because the transition function is not assumed to be total.
So in general not every environment actions can actually occur in a play.
In particular it may happen that a process enters a final state
with no outgoing transition, where
no uncontrollable action can happen.
 
 \paragraph*{Winning games}
 
Our goal is to synthesize  strategies
which ensure that the game terminates and 
all processes are in a final state.

\begin{definition}[Winning strategy]
A strategy $\sigma$ is winning if
the set of $\sigma$-plays  is finite
and in every maximal $\sigma$-play $u$,
every process is in a final state
i.e. $
\forall p \in \PP, \state_p(u)\in  F_p\enspace.
$
\end{definition}
We are interested in the following problem, whose decidability is an open question.
\smallskip

{\noindent \sc Distributed synthesis problem:}
given a distributed game decide
whether  there exists a winning strategy.

\smallskip

If the answer is positive,
the algorithm should as well compute a
winning strategy.

\section{Three decidable classes}
\label{sec:examples}

\paragraph*{Series-parallel games}
A game is \emph{series-parallel} if its dependency alphabet $(A,D)$
is a co-graph i.e. belongs to the smallest class of graphs containing singletons and closed under parallel product and complementation. In this case $A$ has a \emph{decomposition tree},
this is a binary tree whose nodes are subsets of $A$,
its leaves are the singletons $(\{a\})_{a\in A}$,
its root is $A$.
Moreover
every node $B$ with two children $B_0$ and $B_1$
 is the disjoint union of $B_0$ and $B_1$
 and either 
 $B_0\times B_1 \subseteq D$
(serial product)
 or $(B_0\times B_1) \cap D = \emptyset$  (parallel product).

%
The synthesis problem is decidable for series-parallel games~\cite{gastin}.

\paragraph*{Connectedly communicating games}
A game is \emph{$k$-connectedly communicating}  if
for every pair $p,q$ of processes,
 if process $p$ plays $k$ times in parallel of process $q$
 then all further actions of $q$ will be parallel to $p$.
 Formally, for every prime play $uvw$,
$
\label{eq:ccpdef}
(q \not \in\dom(v) \text{ and } |v|_p\geq k )
\implies 
q \not \in \dom(w)\enspace.
$

The MSO theory of the event structure
of a $k$-connectedly communicating game is decidable~\cite{madhu},
which implies that controller synthesis is decidable for theses games.

\paragraph*{Acyclic games}
An acyclic game is a game where processes $\PP$ are the nodes of a tree $T_\PP$
and the domain of every action is a connected set of nodes of $T_\PP$.
The synthesis problem is known to be decidable
for acyclic games such that the domain of each action has size $1$ or $2$~\cite{acyclic}.

\section{Simplifying strategies}
\label{sec:simplifying}

In this section we present an elementary operation called a \emph{shortcut},
which can be used to simplify and reduce the duration of a winning strategy.

To create a shortcut, one selects a $\sigma$-play $xy$
and modify the strategy $\sigma$ so that as soon as any of the processes
sees the play $x$ in its view, this process assumes that not only $x$ but also $xy$ has actually occurred.
In other words, a shortcut is a kind of \emph{cut-and-paste} in the strategy:
we glue on node $x$ the sub-strategy rooted at node $xy$.

The choice of $x$ and $y$ should be carefully performed so that 
the result of the shortcut
is still a strategy.
We provide a sufficient condition for that: $(x,y)$ should be a  \emph{useless repetition}.

The interest of taking shortcuts is the following:
if the original strategy is winning,
then the  strategy obtained by taking the shortcut
is winning as well,
and strictly smaller than the original one.
In the remainder of this section, we formalize these concepts.

\subsection{Locks}

We need to limit the communication between a set of processes,
called a team, and processes outside the team.
This leads to the notion of a $\QQ$-lock:
this is a prime play $u$
such that there is no synchronization  between $\QQ$ and $\PP\setminus \QQ$ in parallel of $u$.

%
%
\begin{definition}
Let $\QQ\subseteq \PP$.
An action $b$ is $\QQ$-safe if $(\dom(b)\subseteq \QQ \text{ or } \dom(b) \cap \QQ=\emptyset)$.
A play $u$ is a \emph{$\QQ$-lock} 
if it is prime and the last action of every prime play parallel to $u$ is $\QQ$-safe.
\end{definition}

The notion of lock is illustrated on
the right handside of Fig.~\ref{fig:example}.
Set $\QQ=\{1,2,3,4,5\}$.
Then
$uv$ is not a $\QQ$-lock because 
$\view_c(uvwc)$ is parallel to $uv$
but $c$ is
not $\QQ$-safe.
Locks occur in a variety of situations,
including the three decidable classes.

\begin{lemma}[Sufficient conditions for $\QQ$-locks]
\label{lem:lock}
Let $u$ be a prime play of a game $G$ and $\QQ\subseteq \PP$.
Each of the following conditions is sufficient for $u$ to be a $\QQ$-lock:
\begin{itemize}
\item[i)]
$\QQ=\PP$.
\item[ii)]
$u$  is a $(\PP\setminus \QQ)$-lock.
\item[iii)]
$\QQ\subseteq\dom(\last(u))$.
\item[iv)]
The game is series-parallel and $\QQ=\dom(B)$
where $B$ is the smallest node of the decomposition tree of $A$ which contains
$\alphabet(u)$.
\item[v)]
The game is connectedly communicating game with bound $k$, $\QQ=\dom(u)$ and $\forall p \in dom(u), |u|_p\geq k$.
\item[vi)]
The game is acyclic with respect to a tree $T_\PP$
and $\QQ$ is the set of descendants in $T_\PP$
of the processes in $\dom(\last(u))$.
\item[vii)]
There are two traces $x$ and $z$ such that $u=xz$ and  $z$ is a $\QQ$-lock
in the game $G_x$ identical to $G$ except  the initial state is changed to $\state(x)$.
\end{itemize}
\end{lemma}

 

\subsection{Taking shortcuts}

%
%
In this section we present a basic operation
used to simplify a strategy, called a \emph{shortcut},
which consists in modifying certain parts of a strategy,
called \emph{useless repetitions}.
These notions rely on the notion of \emph{strategic state} as well as two operations on strategies called \emph{shifting} and \emph{projection}.

\begin{definition}[Residual]
Let $\sigma$ be a strategy,
$u$ a $\sigma$-play and $\QQ\subseteq \PP$.
The $\QQ$-residual of $\sigma$ after $u$ is
the set:
\[
\pi(\sigma,u,\QQ)
=
\{
(v,\sigma(uv))  \mid v \in A_\equiv^*, \dom(v) \subseteq \QQ \text{ and $uv$ is a $\sigma$-play.}
\}\enspace.
\]
%
\end{definition}

A winning strategy may take unnecessarily complicated detours in order to ensure termination.
Such detours are called \emph{useless repetitions}.
\begin{definition}[Strategic state]
Let $\QQ\subseteq \PP$ be a state of processes,
$\sigma$ a strategy and $u$ a prime $\sigma$-play with maximal letter $b$.
The strategic $\QQ$-state of $\sigma$ after $u$ is the tuple
\[
\sstate_{\sigma,\QQ}(u)=
\left(b, \state(u),
\pi\left(\sigma,u, \QQ \setminus \dom(b)\right)
\right)
\enspace.
\]

\end{definition}



\begin{definition}[Useless repetition]
A useless $\QQ$-repetition in a strategy $\sigma$
is a pair of traces $(x,y)$ such that $y$ is not empty,
$xy$ is a $\sigma$-play,  $\dom(y)\subseteq \QQ$,
both $x$ and $xy$ are $\QQ$-locks and $\sstate_{\sigma,\QQ}(x)=\sstate_{\sigma,\QQ}(xy)$.
\end{definition}

The following theorem is the key to our decidability results.

\begin{theorem}
If there exists a winning strategy
then there exists a winning strategy 
without any useless repetition.
\label{theo:uselessdistrib}
\end{theorem}

The proof of this theorem relies on the notion of shortcuts,
an operation which turns a winning strategy into another strategy with strictly shorter duration.

\begin{definition}[Duration of a strategy]
The duration of a strategy $\sigma$ is 
\[
\dur(\sigma)=  \sum_{\text{$u$ maximal $\sigma$-play}} |u|\enspace.
\]
\end{definition}

The duration of a  strategy $\sigma$ may in general be infinite but is finite if $\sigma$ is winning.

\begin{lemma}
\label{lem:shortcut}
Let $(x,y)$ be a  useless $\QQ$-repetition in a  strategy $\sigma$.
Let $\Phi:A^*_\equiv \to A^*_\equiv$ and $\tau$ defined by
$\Phi(u)
=
\begin{cases}
&u \text{ if } x \not \pref u\\
&xyu' \text{ if } x \pref u \text{ and } u=xu'\enspace
\end{cases}$
and
\begin{align*}
\forall p \in \PP, \tau_p(u)
=
\sigma_p(\Phi(\view_p(u))).
\end{align*}
Then $\tau$ is a strategy
called the \emph{$(x,y)$-shortcut of $\sigma$}.
Moreover
for every trace $u$,
\be
\label{eq:tauplay}
(\text{$u$ is a $\tau$-play})
\iff 
(\text{$\Phi(u)$ is a $\sigma$-play})
\enspace.
\ee
If $\sigma$ is a winning strategy then $\tau$ is winning as well and has a strictly smaller duration.
\end{lemma}
\begin{proof}[Sketch of proof of Lemma~\ref{lem:shortcut}]
The full proof can be found in the appendix.
That $\tau$ is a strategy follows from the definition: $\tau_p(u)$ only depends on $\view_p(u)$. To establish~\eqref{eq:tauplay},
the central point is to show that for every $\sigma$-play $xu'$,
\begin{equation*}
\forall p \in\PP,
\sigma_p(\view_p(\Phi(xu'))
=
\sigma_p(\Phi(\view_p(xu')))\enspace.
\end{equation*}
There are three types of plays depending whether:
\begin{enumerate}
\item
$x$ has not occurred  ($x \not \pref u$),
\item
$x$ has occurred in parallel of the process $p$
($x \pref u \land x \not\pref\view_p(u)$),
\item
$p$ knows that
$x$ has occurred
($x \pref \view_p(u)$).
\end{enumerate}
It may happen that $x \pref u$ and
there exists a process $p_2$ in case 2 and a process $p_3$ in case 3. Then process $p_3$ is playing the modified strategy
$xz \to \sigma(xyz)$ while process $p_2$ is still playing the original strategy $\sigma$, which may \emph{a priori} create some $\tau$-plays unrelated with $\sigma$.
The equality of the strategic states in $x$ and $xy$ ensures that the equivalence~\eqref{eq:tauplay} stays valid.

Moreover, thanks to~\eqref{eq:tauplay},
$\dur(\sigma)<\infty$ implies $\dur(\tau)< \dur(\sigma)$ because $y$ is not empty.
And according to~\eqref{eq:tauplay} again,
the set of global states of the maximal plays is the same for $\sigma$ and $\tau$ thus if $\sigma$ is winning then $\tau$ is winning as well.
\end{proof}

\begin{proof}[Proof of Theorem~\ref{theo:uselessdistrib}]
As long as there exists a useless repetition,
take the corresponding shortcut.
According to Lemma~\ref{lem:shortcut},
this creates a sequence $\sigma_0,\sigma_1,\ldots$
of winning strategies whose duration strictly decreases. Thus the sequence is finite and its last element
is a winning strategy without useless repetition.
\end{proof}

\section{Decomposable games}
\label{sec:decomposable}

In this section we introduce \emph{decomposable} games,
for which the \dsp\ is decidable (Theorem~\ref{theo:dec}).
There are actually three notions of decomposability:
structural decomposability, process decomposability and action decomposability.
These three notions form a hierarchy:
structural decomposability implies process decomposability which itself implies action decomposability (Lemma~\ref{lem:hier}).
Known decidable classes are decomposable:
acyclic games are structurally decomposable (Lemma~\ref{lem:acyclic}),
connectedly-communicating games are process decomposable (Lemma~\ref{lem:pdec})
and series-parallel games are action decomposable
(Lemma~\ref{lem:adec}).
Structural decomposability is stable under some  operations between games which leads to new examples
of decidable games (Lemma~\ref{lem:merging}).

\subsection{Decomposability}


The notions of decomposability rely on \emph{preorders} defined on $2^\PP$ or $2^A$. A preorder $\preceq$ is a reflexive and transitive relation.
We denote $\prec$ the relation $(x \prec y) \iff (x \preceq y \land y \not\preceq x)$.
 
\paragraph*{Structural decomposability}
The notion of structural decomposability relies on a preorder
$\preceq$ on $2^\PP$ 
which is 
monotonic with respect to inclusion, i.e. $\forall \QQ,\QQ'\subseteq \PP$,
$
(\QQ \subseteq \QQ' \implies \QQ \preceq \QQ')
$.


\begin{definition}[Structural decomposability]
\label{unif}
A game is \emph{$\preceq$-structurally decomposable}
if for every non-empty prime trace $y\in A^*$
there exists $\QQ\supseteq \dom(y)$
and $b\in\alphabet(y)$ 
such that:
\begin{align*}
&(\QQ \setminus \dom(b)) \prec \QQ\\
&\forall a \in A, (a \ind b \implies \text{$a$ is $\QQ$-safe})\enspace.
\end{align*}
\end{definition}

We have already seen one example of structurally decomposable game.

\begin{lemma}\label{lem:acyclic}
Acyclic games are structurally decomposable.
\end{lemma}
\begin{proof}
Assume the game is acyclic with process tree $T_\PP$.
Set 
$\QQ \preceq \QQ'$ iff 
every process in $\QQ$ has a $T_\PP$-ancestor in $\QQ'$, which is  monotonic with respect to inclusion.
Let $y$ be a prime trace,
$p\in \PP$ the least common ancestor
in $T_\PP$ of processes in $\dom(y)$ and $\QQ$ the set of descendants of $p$.
Then $\dom(y) \subseteq \QQ$.
Moreover, since $y$ is prime and since the domain of every action is a connected subset of $T_\PP$
then $\dom(y)$ is connected as well thus $p\in\dom(y)$
and there exists a letter $b \in\alphabet(y)$ such that $p\in\dom(b)$.
We show that $b$ satisfies the conditions in the definition of structural decomposability.
First, $(\QQ \setminus \dom(b))\preceq \QQ$
and the inequality is strict because 
the only ancestor of $p$ in $\QQ$ is $p$ itself and $p\in\dom(b)$.
Second, let $a \in A$ such that $a\ind b$.
Then $p\not \in \dom(a)$ and since $\dom(a)$ is connected in $T_\PP$,
then either none of the processes in $\dom(a)$ or
all of them are descendants of $p$ in $T_\PP$, i.e. $a$ is $\QQ$-safe.
\end{proof}

\paragraph*{Process decomposability}

The definition of process decomposable games relies on
a parameter $k\in \NN$ and a preorder $\preceq$ on $2^\PP$
which is 
monotonic with respect to inclusion.

\begin{definition}[Process decomposable games]
Fix an integer $k$.
A trace $y$ is \emph{$k$-repeating} if 
\begin{align*}
&y \text{ is not empty and } \forall p\in\dom(y), |y|_p\geq k\enspace.
\end{align*}
A game is \emph{$(\preceq,k)$-process decomposable}
if  for every prime play $xy$,
if $y$ is $k$-repeating then
there exists $\QQ \supseteq \dom(y)$
and a prime prefix $z\pref y$
such that 
$\view_{\last(z)}(xz)$ is a  $\QQ$-lock 
and 
\begin{equation}
\label{pd}
(\QQ \setminus \dom(\last(z))) \prec \dom(y)\enspace.
\end{equation}
\end{definition}

We have already seen one example of process decomposable games.

\begin{lemma}\label{lem:pdec}
Connectedly communicating games
are process decomposable.
\end{lemma}

\paragraph*{Action decomposability}

\newcommand{\FF}{\mathcal{F}}

\newcommand{\dFF}{{\downarrow\FF}}







Action decomposability is defined with respect to a parameter $k\in \NN$
and a preorder $\preceq$ on $2^A$
which is 
monotonic with respect to inclusion.

\begin{definition}[Action decomposable games]
Let $k$ be an integer.
A game is \emph{$(\preceq,k)$ action decomposable}
if  for every prime play $xy$
such that
 $y$ is $k$-repeating,
there exists $\QQ \supseteq \dom(y)$
and a prime prefix $z \pref y$ such that 
$\view_{\last(z)}(xz)$ is a  $\QQ$-lock
and
\[
 \{ a \in A \mid \dom(a) \subseteq (\QQ \setminus \dom(\last(z)) \} \prec \alphabet(y)\enspace.
\]
\end{definition}

We have already seen one example of action decomposable games.

\begin{lemma}\label{lem:adec}
Series-parallel games
are action decomposable.
\end{lemma}

\paragraph*{A hierarchy}

\begin{lemma}\label{lem:hier}
Every structurally decomposable game is process decomposable
and every process decomposable game is action decomposable.
\end{lemma}

Thus \emph{action decomposability} is the most general notion
of decomposability.
In the sequel for the sake of conciseness,
it is simply called \emph{decomposability}.

\subsection{Decidability}
In this section we show that decomposability is a decidable property
and decomposable games have a decidable controller synthesis problem.

\begin{lemma}[Decomposability is decidable]
\label{decdec}
 Whether a game is decomposable is decidable.
There exists a computable function $\decomp$
from games to integers
such that whenever a game $G$
is $(\preceq,k)$ decomposable
 for some $k$, it is $(\preceq,\decomp(G))$ decomposable.
 \end{lemma}
 The proof is elementary and can be found in the appendix.

\begin{theorem}\label{theo:dec}
The distributed synthesis problem is decidable for decomposable games.
\end{theorem}

\begin{proof}[Proof of Theorem~\ref{theo:dec}]
\newcommand{\CC}{\mathcal{C}}

We show that there exists a computable function $f$
from games to integers
such that in every decomposable distributed game $G$
every strategy with no useless repetition
has duration $\leq f(G)$.

Let $\preceq$ be a preorder on $2^A$ compatible with inclusion,
$k'$ an integer and 
$G$ a $(\preceq,k')$ action decomposable distributed game.
Assume $k'=\decomp(G)$ w.l.o.g. (cf. Lemma~\ref{decdec}).

For every set of actions $B\subseteq A$,
denote $G_B$ the game with  actions $B$
and the same processes, initial state and final states than $G$.
The transitions of $G_B$ are all transitions of $G$ whose
 action is in $B$. An action $a \in B$ is controllable in $G_B$ iff it is controllable in $G$.

We show that for every $B\subseteq A$ the game $G_B$ is $(\preceq_B,k')$ decomposable,
where $\preceq_B$ denotes the restriction of $\preceq$ to $2^B$. Let $xy$ be a prime play of $G_B$ 
such that $y$ is $k'$-repeating.
Since  $G$ is $(\preceq,k')$ decomposable, there exists
$\QQ\supseteq \dom(y)$ and a prime prefix $z \pref y$
such that $\view_{\last(z)}(xz)$ is a  $\QQ$-lock in $G$
and $C\prec \alphabet(y)$
where
$
C=  \{ a \in A \mid \dom(a) \subseteq \QQ \text{ and $a \ind \last(z)$} \} $.
Since $\preceq$ is monotonic with respect to inclusion
 then $\{ b \in B \mid \dom(b) \subseteq \QQ \text{ and $b\ind \last(z)$} \}
 =
 (C \cap B) \preceq C \prec \alphabet(y)$
 thus
 $(C \cap B) \prec_B  \alphabet(y)$.
 Since $xy$ is a play in $G_B$ then $\view_{\last(z)}(xz)\pref xy$ is a play in $G_B$ as well.
 And since every play in $G_B$ is a play in $G$, $\view_{\last(z)}(xz)$ is a  $\QQ$-lock not only in $G$ but also in $G_B$.
All conditions of action decomposability are met :
  $G_B$ is $(\preceq_B,k')$ decomposable.

\medskip

Denote $R_B(m)$  the largest size of a complete undirected graph whose edges are labelled with $2^B$ 
and which contains no monochromatic clique of size $\geq m$.
According to Ramsey theorem,
$R_B(m)$ is finite and computable. 
For every $B\subseteq A$,
defined inductively $f(G_B)$ as
:
\[
f(G_B)=
R_{B}\left((k'+|\PP|)\cdot 
|B|\cdot |Q|^{|\PP|} 
\cdot 
2^{ 2^{|A||\PP|\cdot \max\left\{ f\left(G_{B'}\right),{B' \prec B}\right\} }}
\right)
\enspace,
\]
with the convention $\max \emptyset = 0$.

Fix a strategy $\sigma$ with no useless repetition.
We prove that for every  $\sigma$-play $zu$,
\begin{align}
\label{inducdec}
|u| \leq f\left(G_{\alphabet(u)}\right)\enspace.
\end{align}
The proof is by induction on $\alphabet(u)$ with respect to $\preceq$.
The base case when $\alphabet(u)=\emptyset$
is easy, in this case
$|u|=0$.
%

Now let $zu$ be a $\sigma$-play consistent with $\sigma$.
Assume the induction hypothesis holds: for every $\sigma$-play $z'u'$, if
$\alphabet(u') \prec \alphabet(u)$
then
 $|u'| \leq f\left(G_{\alphabet(u')}\right)$. 
 
 We start with computing,
for every non-empty set of letters $B \prec \alphabet(u)$ 
an upper bound on the length of every factorization
$u=u_0u_1\cdots u_Nu_{N+1}$
such that
\begin{align}
&\label{eq:alphabets}
B=
\alphabet(u_1)= \alphabet(u_2)=
\ldots=
\alphabet(u_{N})
\enspace.
\end{align}
For a start, we consider the case where $B$ is connected in the sense
where the dependency graph $D_B=(B, D \cap B\times B)$ is connected.
Set $k= k' + |\PP|$.
For $0 \leq \ell < \frac{N}{k}$,
denote 
$w_\ell$ the concatenation
$w_\ell = u_{1 + \ell k } \cdot u_{2+\ell k} \cdots u_{k + \ell k}$
and 
$h_\ell=zu_0w_1\ldots w_{\ell-1}$.
Let $\RR_B=\dom(B)$ and fix some $c\in B$.

Let 
$0 \leq \ell < \frac{N}{k}$.
We show that $\view_c(w_\ell)$ is $k'$-repeating
and $\view_c(h_\ell w_\ell)=\view_{\RR_B}(h_\ell)\view_c(w_\ell)$.
Since
$
w_\ell = u_{1 + \ell k } \cdot u_{2+\ell k} \cdots u_{k + \ell k}
$,
according to property~\eqref{eq:viewdec}
of views
there exists a sequence 
$\PP\supseteq  \RR_1\supseteq \ldots \supseteq \RR_k$ such that
\begin{align}
\label{eqfacto}
&\view_c(w_\ell)=
\view_{\RR_1}(u_{1 + \ell k })
\view_{\RR_2}(u_{2+\ell k})
\cdots
\view_{\RR_k}(u_{k+\ell k})\enspace
\end{align}
where $\RR_k=\{c\}$ and for every $1\leq i \leq k-1$,
$\RR_i=\RR_{i+1} \cup \dom(
\view_{\RR_{i+1}}(u_{i+1+\ell k})
)$.
Since the sequence
$(\RR_i)_{1 \leq i \leq k'+|\PP|}$ is monotonic,
there exists $i\in k'\dots k'+|\PP|$
such that $\RR_i=\RR_{i+1}$.
Denote $\RR=\RR_i=\RR_{i+1}$
and $B'=\{b \in B, \dom(b) \cap \RR\neq \emptyset\}$
and
$B''=\{b \in B, \dom(b) \subseteq \RR\}$.
By definition of views,
and according to~\eqref{eq:alphabets},
$B'\subseteq \alphabet(\view_{\RR}(u_{i+1+\ell k}))$. Since $\RR=\RR_i=\RR_{i+1}$ and $\RR_i=\RR_{i+1} \cup \dom(
\view_{\RR_{i+1}}(u_{i+1+\ell k})
)$ then
$\dom(
\view_{\RR}(u_{i+1+\ell k})
)\subseteq \RR$
thus $\alphabet(\view_{\RR}(u_{i+1+\ell k}))\subseteq B''$. Since $B'' \subseteq B'$ then finally $B'= \alphabet(\view_{\RR}(u_{i+1+\ell k}))=B''$.
Thus the set $B''$ is a connected component of the graph $D_B=(B,D\cap B \times B)$:
by definition of $B'$ and $B''$, all edges with source $B''$ have target in $B'=B''$.
However by hypothesis $D_B$ is connected thus $B=B'=B''$
and $\RR=\RR_B$.
Finally $\RR_B\subseteq \RR_i\subseteq \RR_1$ and since $\RR_1\subseteq \dom(\view_c(w_\ell))\subseteq \RR_B$,
the sequence $(\RR_i)_{1 \leq i' \leq i}$ is constant equal to $\RR_B$.
%
%
%
Thus, according to~\eqref{eq:alphabets} and the definition of $\RR_B$,
for every $1\leq i' \leq i$,
$\view_{\RR_{i'}}(u_{i'+\ell k})=u_{i'+\ell k}$.
Thus, according to~\eqref{eq:alphabets} and~\eqref{eqfacto}
and since $k' \leq i'$,
every letter of $B$ occurs at least $k'$ times in $\view_c(w_\ell)$
thus
$\view_c(w_\ell)$ is $k'$-repeating and $\view_c(h_\ell w_\ell)=\view_{\RR_B}(h_\ell)\view_c(w_\ell)$.

Since the game is $(\preceq,k')$
decomposable
and $\view_c(w_\ell)$ is $k'$-repeating,
and
$\view_c(h_\ell w_\ell)=\view_{\RR_B}(h_\ell)\view_c(w_\ell)$,
there exists
a superset $\TT^{(\ell)}$ of ${\RR_B}$,
an action $b_\ell$,
and a prime prefix $w'_\ell b_\ell\pref \view_c(w_\ell)$
such that
the play
$z_\ell=\view_{b_\ell}(\view_{\RR_B}(h_\ell) w'_\ell b_\ell)$
is a $\TT^{(\ell)}$-lock
and
\newcommand{\Tl}{\TT^{(\ell)}}
$
B_\ell \prec B \text{ where }
B_\ell =
\{ a \in A \mid \dom(a) \subseteq(\TT^{(\ell)} \setminus \dom(b_\ell)) \}\enspace.
$

For every $0\leq \ell < \frac{N}{k}$,
denote
$\sstate_\ell=\left(
b_\ell, 
\left(s_{\ell,p}\right)_{p\in\PP},
\sigma^{(\ell)}\right)
$
the $\Tl$ strategic state of $\sigma$ after $z_\ell$.
We show two properties of $(\sstate_\ell)_{0 \leq \ell < \frac{N}{k}}$.
\begin{itemize}
\item
First, all elements of $(\sstate_\ell)_{0 \leq \ell < \frac{N}{k}}$ are distinct.
For the sake of contradiction, assume $\sstate_\ell=\sstate_{\ell'}$
for some $0 \leq \ell < \ell' < \frac{N}{k}$.
We show that $z_\ell\sqsubsetneq z_{\ell'}$.
Since $\sstate_\ell=\sstate_{\ell'}$ then  $b_\ell=b_{\ell'}$, denote this letter $b$.
Then
\begin{multline*}
z_\ell
=\view_{b}(\view_{\RR_B}(h_\ell) w'_\ell b)
\pref\view_{b}(\view_{\RR_B}(h_\ell) \view_c(w_\ell))
 = \view_b(\view_c(h_\ell w_\ell))\\
\pref \view_b(\view_c(h_{\ell'} ))
\pref \view_b(\view_{{\RR_B}}(h_{\ell'}))
\sqsubsetneq
\view_{b}(\view_{{\RR_B}}(h_{\ell'}) w'_{\ell'} b)
= z_{\ell'}\enspace,
\end{multline*}
where the second inequality holds
because
$h_\ell w_\ell \pref h_{\ell'} $
since $\ell \leq \ell'-1$,
and the third inequality holds because $c \in B$
thus $\dom(c) \subseteq \RR_B$ hence property~\eqref{viewsub} applies.
Moreover the last inequality is strict because there is at least one more $b$ in
$\view_{b}(\view_{{\RR_B}}(h_{\ell'}) w'_{\ell'} b)$ than in $\view_{b}(\view_{{\RR_B}}(h_{\ell'}))$.
We get a contradiction because by hypothesis there is no useless repetition in $\sigma$,
however, denoting $x=z_\ell$ and $y$ such that $xy=z_{\ell'}$,
the pair $(x,y)$ is a useless $\TT^{(\ell)}$-repetition in $\sigma$:
by hypothesis the strategic $\TT^{(\ell)}$-states of $z_\ell$ and $z_{\ell'}$ are equal and both $x$ and $xy$ are $\TT^{(\ell)}$-locks, moreover $y$ is not empty because $z_\ell\sqsubsetneq z_{\ell'}$
and finally $\dom(y)\subseteq\dom(u_{1+\ell k}\cdots u_{k+ \ell' k}) \subseteq \RR_B \subseteq \TT^{(\ell)}$.
Thus $(x,y)$ is a useless repetition in $\sigma$.
\item
Second, for every $0\leq \ell < \frac{N}{k}$, 
all plays in $\sigma^{(\ell)}=\pi(\sigma,z_\ell,\Tl\setminus \dom(b_\ell))$ have length $\leq m= \max_{ B' \prec B} f\left(G_{B'}\right)$.
Let $z_\ell u'$ be a $\sigma$-play such that $\dom(u')\subseteq (\Tl\setminus \dom(b_\ell))$.
Then $\alphabet(u') \subseteq B_\ell$.
Since $\preceq$ is monotonic with respect to inclusion,
$
\alphabet(u') \preceq B_\ell\prec B \preceq \alphabet(u).
$
Thus by induction hypothesis,
$|u'| \leq f\left(G_{\alphabet(u')}\right)\leq m$.
\end{itemize}

According to the second property,
there
are at most $2^{ 2^{m|A||\PP|}}$ different residuals appearing in the sequence
$(\sigma^{(\ell)})_{0\leq \ell < \frac{N}{k}}$.
Thus the sequence $(\sstate_\ell)_{0 \leq \ell< \frac{N}{k}}$ takes at most 
$
K=
|B|\cdot |Q|^{|\PP|} 
\cdot 2^{ 2^{m|A||\PP| }}\enspace
$
different values.
And according to the first property, all these states are different thus 
$N \leq k\cdot K
$.

The inequality $N \leq k\cdot K$ has been established  under the assumption that $D_B$ is connected.
The general case reduces to this case: let $C$
be a connected component of $D_B$ and for $1\leq i \leq N$
let $v_i$ be the projection of $u_i$ on $C$.
Then $\forall 1\leq i \leq N, \alphabet(v_i)=C$
and
there exists $u'_0$ such that
$u=u'_0 v_1v_2\ldots v_N u_{N+1}$ thus $N \leq k\cdot K$.

Let us reformulate the inequality $N \leq k\cdot K$ as a property of an undirected complete graph
with edges colored by $2^A$.
Let $u=a_1a_2\cdots a_{|u|}$ the factorization of $u$ into its letters.
Let $J_u$ be the complete graph with vertices $1,\ldots, |u|$
and the label of the edge $\{i , j\}$ with $i<j$ is the set of letters $\{a_i,\ldots ,a_j\}$.
Then every  monochromatic clique of $J_u$ has size $\leq k\cdot K$.
Thus, according to Ramsey theorem,
$
|u|
\leq
R_\TT\left(k\cdot K\right)
=
R_\TT\left((k'+|\PP|)\cdot K\right)
$,
which completes the inductive step.

As a consequence, winning strategies in $G$ can be looked for in the finite family of strategies
whose all plays have length $\leq f(G)$
with $f(G)$ computable.
As a consequence,
the synthesis problem
can be solved
by enumerating all these strategies and testing whether any of them is winning.
For testing whether a strategy of finite duration
is winning the algorithm simply checks that the global state of all the maximal plays is final.
\end{proof}

\subsection{New examples of decidable games}

The three classes of games whose decidability is already known
are decomposable  (cf Lemmas~\ref{lem:acyclic},~\ref{lem:pdec} and~\ref{lem:adec}).
In this section we give some new examples of decidable games.

\begin{lemma}\label{lem:fourp}
Four players games are  structurally decomposable.
\end{lemma}

Although our techniques do not seem to provide an algorithm for solving games with five processes, they can address a special case of those.

\begin{lemma}\label{lem:fivep}
Let $G$ be a distributed game with five processes.
Assume that the number of actions
that a process can successively play in a row without 
synchronizing simultaneously with two other processes
is bounded.
Then $G$ is process decomposable.
\end{lemma}

Another decidable example is the class of \emph{majority games}:
\begin{lemma}[Majority games]\label{lem:majority}
Assume that every non-local action synchronizes a majority of the processes i.e. for every action $a$,
$
|\dom(a)| = 1 \text{ or }   | \dom(a) |  \geq | \PP \setminus \dom(a) |.
$
Then the game is  structurally decomposable.
\end{lemma}

The class of structurally decomposable games
is stable under projection and merge.

\begin{definition}[Projecting games]
Let $G$ be a game with processes $\PP$ and
alphabet $(A_p)_{p\in\PP}$.
Let 
$\PP'\subseteq \PP$
a subset of the processes.
The projection of $G$ on $\PP'$ is the game $G'$
with processes $\PP'$ and alphabet
$A' = \{ a \in A \mid \dom(a) \cap \PP' \neq \emptyset \}$
partitioned in $(A' \cap A_p)_{p\in\PP'}$.
The states of a process $p\in\PP'$ are the same in $G$ and $G'$, every transition 
$\delta \in \{a\} \times \Pi_{p\in\dom(a)} Q_p\times Q_p$
of $G$ on a letter $a\in A'$
is projected to $\{a\} \times \Pi_{p\in\dom(a)\cap \PP'} Q_p\times Q_p$,
and every transition on a letter $a \not \in A'$ is simply deleted.
\end{definition}

The following result combines two structurally decomposable games into one.

\begin{lemma}[Merging games]\label{lem:merging}
Let $G$ be a game,
and $\PP_0, \PP_1\subseteq \PP$
 two set of processes such that $\PP=\PP_0\cup\PP_1$ and for every action $a\in A$,
\[
(\dom(a) \cap \PP_0\neq \emptyset)\land (\dom(a) \cap \PP_1\neq \emptyset)
\implies(\PP_0\cap \PP_1\subseteq \dom(a))\enspace.
\]
If both projections of $G$ on $(\PP_0\setminus \PP_1)$ and $(\PP_1\setminus \PP_0)$ are structurally decomposable then $G$ is structurally decomposable.
\end{lemma}

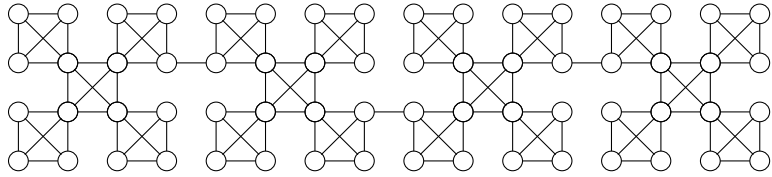
\begin{figure}[ht]
\begin{center}
\begin{tikzpicture}
\begin{scope}[scale=1.3]

\begin{scope}
\draw (0,0) circle (0.1);
\draw (0.5,0) circle (0.1);
\draw (0.5,0.5) circle (0.1);
\draw (0,0.5) circle (0.1);
\draw (0.1,0) -- (0.4,0) ;
\draw (0,0.1) -- (0,0.4) ;
\draw (0.1,0.5) -- (0.4,0.5) ;
\draw (0.5,0.4) -- (0.5,0.1) ;
\draw (0.065,0.065) -- (0.435,0.435) ;
\draw (0.065,0.435) -- (0.435,0.065) ;
\end{scope}

\begin{scope}[shift={(0.5,0.5)}]
\draw (0,0) circle (0.1);
\draw (0.5,0) circle (0.1);
\draw (0.5,0.5) circle (0.1);
\draw (0,0.5) circle (0.1);
\draw (0.1,0) -- (0.4,0) ;
\draw (0,0.1) -- (0,0.4) ;
\draw (0.1,0.5) -- (0.4,0.5) ;
\draw (0.5,0.4) -- (0.5,0.1) ;
\draw (0.065,0.065) -- (0.435,0.435) ;
\draw (0.065,0.435) -- (0.435,0.065) ;
\end{scope}

\begin{scope}[shift={(-0.5,0.5)}]
\draw (0,0) circle (0.1);
\draw (0.5,0) circle (0.1);
\draw (0.5,0.5) circle (0.1);
\draw (0,0.5) circle (0.1);
\draw (0.1,0) -- (0.4,0) ;
\draw (0,0.1) -- (0,0.4) ;
\draw (0.1,0.5) -- (0.4,0.5) ;
\draw (0.5,0.4) -- (0.5,0.1) ;
\draw (0.065,0.065) -- (0.435,0.435) ;
\draw (0.065,0.435) -- (0.435,0.065) ;
\end{scope}

\begin{scope}[shift={(-0.5,-0.5)}]
\draw (0,0) circle (0.1);
\draw (0.5,0) circle (0.1);
\draw (0.5,0.5) circle (0.1);
\draw (0,0.5) circle (0.1);
\draw (0.1,0) -- (0.4,0) ;
\draw (0,0.1) -- (0,0.4) ;
\draw (0.1,0.5) -- (0.4,0.5) ;
\draw (0.5,0.4) -- (0.5,0.1) ;
\draw (0.065,0.065) -- (0.435,0.435) ;
\draw (0.065,0.435) -- (0.435,0.065) ;
\end{scope}

\begin{scope}[shift={(0.5,-0.5)}]
\draw (0,0) circle (0.1);
\draw (0.5,0) circle (0.1);
\draw (0.5,0.5) circle (0.1);
\draw (0,0.5) circle (0.1);
\draw (0.1,0) -- (0.4,0) ;
\draw (0,0.1) -- (0,0.4) ;
\draw (0.1,0.5) -- (0.4,0.5) ;
\draw (0.5,0.4) -- (0.5,0.1) ;
\draw (0.065,0.065) -- (0.435,0.435) ;
\draw (0.065,0.435) -- (0.435,0.065) ;
\end{scope}

\begin{scope}[shift={(2,0)}]

\draw (-0.9,0) -- (-0.6,0) ;

\begin{scope}
\draw (0,0) circle (0.1);
\draw (0.5,0) circle (0.1);
\draw (0.5,0.5) circle (0.1);
\draw (0,0.5) circle (0.1);
\draw (0.1,0) -- (0.4,0) ;
\draw (0,0.1) -- (0,0.4) ;
\draw (0.1,0.5) -- (0.4,0.5) ;
\draw (0.5,0.4) -- (0.5,0.1) ;
\draw (0.065,0.065) -- (0.435,0.435) ;
\draw (0.065,0.435) -- (0.435,0.065) ;
\end{scope}

\begin{scope}[shift={(0.5,0.5)}]
\draw (0,0) circle (0.1);
\draw (0.5,0) circle (0.1);
\draw (0.5,0.5) circle (0.1);
\draw (0,0.5) circle (0.1);
\draw (0.1,0) -- (0.4,0) ;
\draw (0,0.1) -- (0,0.4) ;
\draw (0.1,0.5) -- (0.4,0.5) ;
\draw (0.5,0.4) -- (0.5,0.1) ;
\draw (0.065,0.065) -- (0.435,0.435) ;
\draw (0.065,0.435) -- (0.435,0.065) ;
\end{scope}

\begin{scope}[shift={(-0.5,0.5)}]
\draw (0,0) circle (0.1);
\draw (0.5,0) circle (0.1);
\draw (0.5,0.5) circle (0.1);
\draw (0,0.5) circle (0.1);
\draw (0.1,0) -- (0.4,0) ;
\draw (0,0.1) -- (0,0.4) ;
\draw (0.1,0.5) -- (0.4,0.5) ;
\draw (0.5,0.4) -- (0.5,0.1) ;
\draw (0.065,0.065) -- (0.435,0.435) ;
\draw (0.065,0.435) -- (0.435,0.065) ;
\end{scope}

\begin{scope}[shift={(-0.5,-0.5)}]
\draw (0,0) circle (0.1);
\draw (0.5,0) circle (0.1);
\draw (0.5,0.5) circle (0.1);
\draw (0,0.5) circle (0.1);
\draw (0.1,0) -- (0.4,0) ;
\draw (0,0.1) -- (0,0.4) ;
\draw (0.1,0.5) -- (0.4,0.5) ;
\draw (0.5,0.4) -- (0.5,0.1) ;
\draw (0.065,0.065) -- (0.435,0.435) ;
\draw (0.065,0.435) -- (0.435,0.065) ;
\end{scope}

\begin{scope}[shift={(0.5,-0.5)}]
\draw (0,0) circle (0.1);
\draw (0.5,0) circle (0.1);
\draw (0.5,0.5) circle (0.1);
\draw (0,0.5) circle (0.1);
\draw (0.1,0) -- (0.4,0) ;
\draw (0,0.1) -- (0,0.4) ;
\draw (0.1,0.5) -- (0.4,0.5) ;
\draw (0.5,0.4) -- (0.5,0.1) ;
\draw (0.065,0.065) -- (0.435,0.435) ;
\draw (0.065,0.435) -- (0.435,0.065) ;
\end{scope}

\end{scope}


\begin{scope}[shift={(4,0)}]

\draw (-0.9,0.5) -- (-0.6,0.5) ;

\begin{scope}
\draw (0,0) circle (0.1);
\draw (0.5,0) circle (0.1);
\draw (0.5,0.5) circle (0.1);
\draw (0,0.5) circle (0.1);
\draw (0.1,0) -- (0.4,0) ;
\draw (0,0.1) -- (0,0.4) ;
\draw (0.1,0.5) -- (0.4,0.5) ;
\draw (0.5,0.4) -- (0.5,0.1) ;
\draw (0.065,0.065) -- (0.435,0.435) ;
\draw (0.065,0.435) -- (0.435,0.065) ;
\end{scope}

\begin{scope}[shift={(0.5,0.5)}]
\draw (0,0) circle (0.1);
\draw (0.5,0) circle (0.1);
\draw (0.5,0.5) circle (0.1);
\draw (0,0.5) circle (0.1);
\draw (0.1,0) -- (0.4,0) ;
\draw (0,0.1) -- (0,0.4) ;
\draw (0.1,0.5) -- (0.4,0.5) ;
\draw (0.5,0.4) -- (0.5,0.1) ;
\draw (0.065,0.065) -- (0.435,0.435) ;
\draw (0.065,0.435) -- (0.435,0.065) ;
\end{scope}

\begin{scope}[shift={(-0.5,0.5)}]
\draw (0,0) circle (0.1);
\draw (0.5,0) circle (0.1);
\draw (0.5,0.5) circle (0.1);
\draw (0,0.5) circle (0.1);
\draw (0.1,0) -- (0.4,0) ;
\draw (0,0.1) -- (0,0.4) ;
\draw (0.1,0.5) -- (0.4,0.5) ;
\draw (0.5,0.4) -- (0.5,0.1) ;
\draw (0.065,0.065) -- (0.435,0.435) ;
\draw (0.065,0.435) -- (0.435,0.065) ;
\end{scope}

\begin{scope}[shift={(-0.5,-0.5)}]
\draw (0,0) circle (0.1);
\draw (0.5,0) circle (0.1);
\draw (0.5,0.5) circle (0.1);
\draw (0,0.5) circle (0.1);
\draw (0.1,0) -- (0.4,0) ;
\draw (0,0.1) -- (0,0.4) ;
\draw (0.1,0.5) -- (0.4,0.5) ;
\draw (0.5,0.4) -- (0.5,0.1) ;
\draw (0.065,0.065) -- (0.435,0.435) ;
\draw (0.065,0.435) -- (0.435,0.065) ;
\end{scope}

\begin{scope}[shift={(0.5,-0.5)}]
\draw (0,0) circle (0.1);
\draw (0.5,0) circle (0.1);
\draw (0.5,0.5) circle (0.1);
\draw (0,0.5) circle (0.1);
\draw (0.1,0) -- (0.4,0) ;
\draw (0,0.1) -- (0,0.4) ;
\draw (0.1,0.5) -- (0.4,0.5) ;
\draw (0.5,0.4) -- (0.5,0.1) ;
\draw (0.065,0.065) -- (0.435,0.435) ;
\draw (0.065,0.435) -- (0.435,0.065) ;
\end{scope}

\end{scope}

\begin{scope}[shift={(-2,0)}]

\draw (1.1,0.5) -- (1.4,0.5) ;

\begin{scope}
\draw (0,0) circle (0.1);
\draw (0.5,0) circle (0.1);
\draw (0.5,0.5) circle (0.1);
\draw (0,0.5) circle (0.1);
\draw (0.1,0) -- (0.4,0) ;
\draw (0,0.1) -- (0,0.4) ;
\draw (0.1,0.5) -- (0.4,0.5) ;
\draw (0.5,0.4) -- (0.5,0.1) ;
\draw (0.065,0.065) -- (0.435,0.435) ;
\draw (0.065,0.435) -- (0.435,0.065) ;
\end{scope}

\begin{scope}[shift={(0.5,0.5)}]
\draw (0,0) circle (0.1);
\draw (0.5,0) circle (0.1);
\draw (0.5,0.5) circle (0.1);
\draw (0,0.5) circle (0.1);
\draw (0.1,0) -- (0.4,0) ;
\draw (0,0.1) -- (0,0.4) ;
\draw (0.1,0.5) -- (0.4,0.5) ;
\draw (0.5,0.4) -- (0.5,0.1) ;
\draw (0.065,0.065) -- (0.435,0.435) ;
\draw (0.065,0.435) -- (0.435,0.065) ;
\end{scope}

\begin{scope}[shift={(-0.5,0.5)}]
\draw (0,0) circle (0.1);
\draw (0.5,0) circle (0.1);
\draw (0.5,0.5) circle (0.1);
\draw (0,0.5) circle (0.1);
\draw (0.1,0) -- (0.4,0) ;
\draw (0,0.1) -- (0,0.4) ;
\draw (0.1,0.5) -- (0.4,0.5) ;
\draw (0.5,0.4) -- (0.5,0.1) ;
\draw (0.065,0.065) -- (0.435,0.435) ;
\draw (0.065,0.435) -- (0.435,0.065) ;
\end{scope}

\begin{scope}[shift={(-0.5,-0.5)}]
\draw (0,0) circle (0.1);
\draw (0.5,0) circle (0.1);
\draw (0.5,0.5) circle (0.1);
\draw (0,0.5) circle (0.1);
\draw (0.1,0) -- (0.4,0) ;
\draw (0,0.1) -- (0,0.4) ;
\draw (0.1,0.5) -- (0.4,0.5) ;
\draw (0.5,0.4) -- (0.5,0.1) ;
\draw (0.065,0.065) -- (0.435,0.435) ;
\draw (0.065,0.435) -- (0.435,0.065) ;
\end{scope}

\begin{scope}[shift={(0.5,-0.5)}]
\draw (0,0) circle (0.1);
\draw (0.5,0) circle (0.1);
\draw (0.5,0.5) circle (0.1);
\draw (0,0.5) circle (0.1);
\draw (0.1,0) -- (0.4,0) ;
\draw (0,0.1) -- (0,0.4) ;
\draw (0.1,0.5) -- (0.4,0.5) ;
\draw (0.5,0.4) -- (0.5,0.1) ;
\draw (0.065,0.065) -- (0.435,0.435) ;
\draw (0.065,0.435) -- (0.435,0.065) ;
\end{scope}

\end{scope}

\end{scope}

\end{tikzpicture}
\end{center}

\caption{\label{fig:network} A decidable process architecture.}
\end{figure}

The merge operation can combine two structurally decomposable games in order to create a new one.
For example all acyclic games can be obtained this way,
since $3$-player games are structurally decomposable
and every tree with more than three 
nodes
can be obtained by merging two strictly smaller subtrees.
This technique can go beyond acyclic games,
by merging together 4-player games
and majority games.
The graph of processes is an undirected graph with nodes $\PP$
 and there is an edge between $p$ and $q$
whenever both $p$ and $q$ both belong to the domain of one of the actions.
Then all the games whose graph of processes
is contained in the one depicted on Fig.~\ref{fig:network} are structurally decomposable.

\section*{Conclusion}
We considered the \dsp,
which aims at controlling asynchronous automata
using automatically 
synthesized controllers with causal memory.
We presented a theorem that unifies several known decidability results and provide new ones.

The decidability of this problem is still open to our knowledge, even in the simple case where the graph of processes is a ring of five processes where each process can interact only with both its  neighbors.

Another intriguing open problem is the case of \emph{weakly $k$-connectedly communicating} plants.
In such a plant, whenever two processes play both $k$ times in a row without hearing from each other,
they will never hear from each other anymore. 
It is not known whether the MSO theory of
the corresponding event structures is decidable or not~\cite{madhu},
and we do not know either how to use techniques of this paper to solve this class of games.

\section*{Acknowledgements}

We thank Blaise Genest, Anca Muscholl, Igor Walukiewicz, Paul Gastin and Marc Zeitoun for interesting discussions on the topic.
Moreover we thank one of the reviewers
of a previous version, who spotted several mistakes and did provide very useful comments which led to several improvements in the presentation of the results.

\newpage

%
%

\section*{Appendix}

\section{Definition of the $\QQ$-view}

For every set of processes $\QQ$ and \emph{word} $u\in A^*$,
we define inductively the $\QQ$-view of $u$ as
follows.
If $u$ is empty, its view is empty.
If $u$ is a word and $a$ is a letter then:
\begin{equation}\label{defview}
\view_\QQ(ua)
=
\begin{cases}
\view_\QQ(u)&  \text{ if }  \dom(a) \cap \QQ=\emptyset\\
\view_{\QQ\cup \dom(a)}(u)a
&  \text{ if }  \dom(a) \cap \QQ\neq\emptyset\enspace.
\end{cases}
\end{equation}

An easy induction shows that for every words $u,v$,
\begin{equation}
\label{eq:viewdecapp}
\view_\QQ(uv)=\view_{\QQ'}(u)\view_\QQ(v)
\text{ where }
\QQ'=\QQ\cup \dom(\view_\QQ(v))\enspace.
\end{equation}

\begin{lemma}\label{lem:vi}
Let $\QQ$ be a set of processes and $u,v$ two words and $a,b$ two letters such that $a \ind b$,
$
\view_\QQ(uabv)=\view_\QQ(ubav)\enspace.
$
\end{lemma}
\begin{proof}
According to~\eqref{eq:viewdecapp},
\[
\view_\QQ(uabv)
=
\view_{\QQ''}(u)
\view_{\QQ'}(ab)
\view_\QQ(v)
\]
with $\QQ'= \QQ \cup \dom(\view_\QQ(v))$
and $\QQ''= \QQ'\cup \dom(\view_{\QQ'}(ab))$.

Then $\view_{\QQ'}(ab)=\view_{\QQ'''}(a)\view_{\QQ'}(b)$
where $\QQ'''=\QQ'\cup\dom(\view_{\QQ'''}(a))$. However $a\ind b$ thus 
$\dom(\view_{\QQ'''}(a))\cap \dom(b)=\emptyset$ hence $\view_{\QQ'''}(a)=\view_{\QQ'}(a)$ thus
$\view_{\QQ'}(ab)=\view_{\QQ'}(ba)$
and by symetry
$\view_\QQ(uabv)=\view_\QQ(ubav)$.
\end{proof}

According to Lemma~\ref{lem:vi},
the view is independent by commutation of independent letters, thus its definition extends to traces.

A simple induction provides several useful properties of views.
\begin{align}
\label{viewpref}
&\view_\QQ(u) \pref u\\
\label{viewall}
&\view_{\dom(u)}(u)=u
\\
\label{viewempty}
& (\view_\QQ(u)=\epsilon) \iff (\dom(u)\cap \QQ=\emptyset)
\\
\label{viewsubapp}
& (\QQ\subseteq \QQ') \implies 
(\view_\QQ(u)
 \pref
\view_{\QQ'}(u))\\
\label{eq:viewview2}
&\text{$\view_\QQ(\view_\QQ(u)) = \view_\QQ(u)$}\\
\label{eq:viewview}
&\text{$\view_\QQ(uv) = \view_\QQ(u\view_\QQ(v))$}
\enspace.
\end{align}

To establish that the definition of views given in Definition~\ref{def:views} is equivalent
to the one given by~\eqref{defview},
we have to show:
\begin{lemma}
For every set of processes $\QQ$,
every trace $u$ has a longest suffix $v$ such that $\dom(v)\cap \QQ=\emptyset$.
And
$u=\view_\QQ(u)v$.
\end{lemma}
\begin{proof}
According to~\eqref{viewpref},
there exists $w$ such that $u=\view_\QQ(u)w$.
We show that
$\dom(w)\cap \QQ=\emptyset$.
According to~\eqref{eq:viewdec},
$\view_\QQ(u)=\view_{\QQ'}(u)\view_\QQ(w)$
with $\QQ'=\QQ\cup \dom(\view_\QQ(w))$.
Since $\QQ'\supseteq \QQ$
then
$|\view_\QQ(u)| \leq |\view_{\QQ'}(u)|$
according to~\eqref{viewsubapp}
thus $|\view_\QQ(w)|=0$ hence
$\dom(w)\cap \QQ=\emptyset$ according to~\eqref{viewempty}.
Let $v$ be a suffix of $u$
such that $\dom(v)\cap \QQ=\emptyset$.
Let $u'$ such that $u=u'v$. 
Then $\view_\QQ(u)=\view_\QQ(u')$
thus $\view_\QQ(u)\pref u'$ hence $|u'| \geq |u|-|w|$ hence $|v|\leq |w|$.
\end{proof}

\section{Elementary properties of traces}

Not all properties of the concatenation operator
and the prefix relation on words are preserved on traces,
however the following are:
\begin{align}
\label{eq:prefantisym}
&\forall u,v\in A^*, ((u\pref v) \land (v\pref u) \implies u=v)\enspace,\\
\label{eq:prefcancel}
&\forall u,v,w\in A^*, (uv=uw) \implies (v=w)\enspace,\\
&
\label{eq:preftrace}
\forall u,v,w\in A^*, (uv\pref uw)\implies(v\pref w)\enspace.
\end{align}

The following two lemmas list some basic properties of traces used in the proofs.


\begin{lemma}\label{lem:facto}
Let $u,v,x,y$ some traces such that
$uv=xy$.
Then there exists factorizations
$x=x'x''$ and $y=y'y''$ such that:
\begin{align}
&u=x'y'\\
&v=x''y''\\
&\dom(x'') \cap \dom(y')=\emptyset\enspace.
\end{align}
\end{lemma}
\begin{proof}
By induction on $|uv|$.
The case where $uv$ is empty is trivial.
Otherwise let $a$ be a maximal action of $v$ so that
$v=v_0a$. There are two cases.

If $\dom(a) \cap \dom(y)\neq\emptyset$
then since $a$ is a maximal action of $xy$
and does not commute with $y$,
$a$ is a maximal action of $y$.
Thus $y$ factorizes as $y=y_0a$
and we apply the induction hypothesis to the equality
$uv_0=xy_0$ and append $a$ to $y_0''$.

If $\dom(a) \cap \dom(y)=\emptyset$
then $a$ is a maximal action of $x$ which factorizes as $x=x_0a$. We apply the induction hypothesis
to the equality $uv_0=x_0y$ and append $a$ to $x_0''$.
\end{proof}

We define the notion of $\QQ$-prime trace.
\begin{definition}[$\QQ$-prime trace]
A trace is $\QQ$-prime if $\view_\QQ(u)=u$.
\end{definition}

We make use of the following properties of traces.

\begin{lemma}\label{lem:props}
For every trace $u,v,x\in A^*$ and $a\in A$ and $B\subseteq A$,\begin{align}
&\label{eq:primsuff}
\text{$uv$ is $\QQ$-prime $\implies v$ is $\QQ$-prime}\\
&\label{eq:primconcat2}
\text{$u$ and $v$ are $\QQ$-prime $\implies uv$ is $\QQ$-prime}\\
&\label{eq:primsuff2}
\text{If } ua  \text{ is prime},
 (av \text{ is $\QQ$-prime } \iff  uav \text{ is $\QQ$-prime })\\
%
%
\label{eq:primsuff4}
&(u\pref \view_\QQ(uv))\iff (\view_\QQ(uv) = u\view_\QQ(v))\\
%
&
\notag
\text{If } ua  \text{ is prime},\\
&
\hspace{.5cm}
\label{eq:primconcat}
\view_\QQ(uav) = ua\view_\QQ(v)  \iff  \view_\QQ(av) = a \view_\QQ(v)\\
&\label{eq:viewempty}
(\view_\QQ(uv)= \view_\QQ(u))\implies \view_\QQ(v)=\epsilon\\
& \label{comm}
\left((a \pref u) \land (x \pref u) \land (a\not \pref x) \right)
\implies
\left( (\dom(a)\cap \dom(x)=\emptyset) \land (ax \pref u)\right)\enspace.
\end{align}

\end{lemma}
\begin{proof}
We prove~\eqref{eq:primsuff}.
If the last letter of a word $v'\in v$
is not in $\QQ$, then the same holds for every $u'v'$ where $u'$ is a linearization of the trace $u$ thus $uv$ is not $\QQ$-prime since $u'v'$ is a linearization of the trace $uv$.

We prove~\eqref{eq:primconcat2}.
Assume both $u$ and $v$ are $\QQ$-prime.
Every linearization of $uv$ is an interleaving of a linearization
of $u$ and a linearization of $v$ thus it terminates with a letter
whose domain intersects $\QQ$. Hence $uv$ is $\QQ$-prime.

We prove~\eqref{eq:primsuff2}.
Assume $ua$ prime.
The converse implication follows from~\eqref{eq:primsuff}.
Assume $av$ is $\QQ$-prime.
We prove that $uav$ is $\QQ$-prime by induction on $|u|$.
If $|u|=0$ then $u=\epsilon$ and $uav=av$ is $\QQ$-prime by hypothesis.
By induction let $n\in \NN$ and assume $u'av$ is $\QQ$-prime
for all $u'$ such that $|u'|\leq n$. Let $u$ such that $|u|=n+1$,
we prove that $uav$ is $\QQ$-prime.
Since $|u|=n+1$,
there exists $b\in A$ and $u'\in A^*$
such that $u=bu'$ and $|u'|=n$.
Using~\eqref{eq:primsuff} and the induction hypothesis
so on one hand we know that $u'av$ is $\QQ$-prime.
By definition of a trace, for any trace $w$,
\be
\label{eq:commute}
bw =\{ xbz \mid x,z \text{ words on $A$ }, xz \in w, b\ind x\}\enspace.
\ee
Let $y$ a linearization of $uav=bu'av$,
we prove that the last letter of $y$ is in $A_\QQ=\cup_{p\in\QQ}A_p$.
According to~\eqref{eq:commute},
$y$ factorizes as $y=xbz$ with $xz\in u'av$ and $x\ind b$.
Since $xz\in u'av$ and $u'av$ is $\QQ$-prime,
if $z$ is not empty then it ends with a letter in $\cup_{p\in\QQ}A_p$ and so does $y$.
Assume now that $z$ is empty, then $y=xb$ with $x\in u'av$ and $x\ind b$.
Since $y\in bu'av$ then $\alphabet(u'a)\subseteq \alphabet(y)$
and $\alphabet(v)\subseteq \alphabet(y)$.
Since $\alphabet(y)=\alphabet(x)\cup \{b\}$
and $x\ind b$ every letter of $u'a$ and $av$ commute with $b$
thus $bu'a=u'ab$ and $bv = vb$.
Since $bu'a=ua$ is prime, $bu'a=u'ab$ implies $a=b$.
Since $bv=vb$ then $av=va$ and since $av$ is $\QQ$-prime, $a=b\in B$.
Finally $b\in A_\QQ$ and since $y=xb$ the last letter of $y$ is in $A_\QQ$,
which terminates the proof of the inductive step,
and the proof of~\eqref{eq:primsuff2}.

We prove~\eqref{eq:primsuff4}.
The converse implication in~\eqref{eq:primsuff4} is obvious so it is enough 
to prove the direct implication.
Assume $u\pref \view_\QQ(uv)$.
According to~\eqref{eq:prefantisym}
it is enough to prove
both
$\view_\QQ(uv) \pref u\view_\QQ(v)$
and
$u\view_\QQ(v)\pref \view_\QQ(uv)$.
We start with $u\view_\QQ(v)\pref \view_\QQ(uv)$.
Since $u\pref \view_\QQ(uv)$, then $\view_\QQ(uv)=uw$ for some $w\in A^*$
and $uv=uww'$ for some $w'$ such that $\dom(w')\cap\QQ=\emptyset$.
Then $v=ww'$ according to~\eqref{eq:prefcancel}
and since $\dom(w')\cap \QQ=\emptyset$, then $\view_\QQ(v)\pref w$,
thus $u\view_\QQ(v)\pref uw=\view_\QQ(uv)$ and we got the first prefix relation.
Now we prove the converse prefix relation.
Since $\view_\QQ(v)\pref w$ then by definition of $\view_\QQ$ there exists $w''\in A^*$
such that $w=\view_\QQ(v) w''$ and $\dom(w'')\cap \QQ=\emptyset$.
Then $uv = u\view_\QQ(v) w''w'$ and $\dom(w''w')\cap\QQ=\emptyset$
thus by definition of $\view_\QQ$,
$\view_\QQ(uv) \pref u\view_\QQ(v)$.
By definition of $w$ this implies
$uw \pref u\view_\QQ(v)$ thus according to~\eqref{eq:preftrace}
$w\pref \view_\QQ(v)$. Finally $w=\view_\QQ(v)$
and $u\view_\QQ(v)=uw=u\view_\QQ(v)$
which terminates the proof of~\eqref{eq:primsuff4}.


By definition $\view_\QQ(uv)$ is the shortest prefix of $uv$
such that $uv$=$\view_\QQ(uv)v'$ with 
$\dom(v')\cap \QQ=\emptyset$,
thus by hypothesis there exists $w'$ such that $uv=uww'v'$.
$v=ww'v'$ thus by definition of $\view_\QQ(v)$ again,
$ww'\pref \view_\QQ(v)$ thus $w\pref\view_\QQ(v)$.

We prove~\eqref{eq:primconcat}.
Let $\QQ'=\QQ \cup \dom(\view_\QQ(v))$
Then according to~\eqref{eq:viewdec}, 
$\view_\QQ(uav)  = \view_{\QQ'}(ua) \view_\QQ(v)$
and
$\view_\QQ(av)  = \view_{\QQ'}(a) \view_\QQ(v)$
thus 
$(\view_\QQ(av)=a \view_\QQ(v)) \iff (\dom(a)\cap \dom(\QQ')\neq \emptyset) \iff (\view_{\QQ'}(ua)=ua)$ (since $ua$ is prime).

We prove~\eqref{eq:viewempty}.
Let $\QQ'=\QQ\cup \dom(\view_\QQ(v))$.
Then according to~\eqref{eq:viewdec},
$\view_\QQ(uv)=\view_{\QQ'}(u)\view_\QQ(v)$.
Since $\QQ\subseteq {\QQ'}$ then
$\view_\QQ(u)\pref \view_{\QQ'}(u)$
thus $\view_\QQ(uv)=\view_\QQ(u)$
implies $\view_\QQ(u)=\view_{\QQ'}(u)=\view_{\QQ'}(u)\view_\QQ(v)$
hence $\view_\QQ(v)= \emptyset$.

Finally, we prove~\eqref{comm}.
Assume $a \pref u$ and $x \pref u$
and $a\not \pref x$.
We show that $\dom(a)\cap \dom(x)=\emptyset$
and $ax \pref u$.
Let $u_1,u_2$ such that
$u=au_1$ and $u=xu_2$.
Then according to Lemma~\ref{lem:facto},
there exist factorizations $a=a'a''$ and
$u_1=u_3u_4$ such that $x=a'u_3$
and $u_2=a''u_4$ and $\dom(a'')\cap \dom(u_3)=\emptyset$.
Since $a$ is a letter, either 
$(a'=a \land a''=\epsilon)$
or
$(a'=\epsilon \land a''=a)$.
However $a \not \pref x$
thus $a'\neq a$ hence
$(a'=\epsilon \land a''=a)$.
Thus $\dom(a)\cap \dom(x)=\dom(u_3)\cap \dom(x)=\emptyset$. And $u=xu_2=xau_4=axu_4$ thus $ax \pref u$.
\end{proof}


\begin{lemma}\label{lem:lcp}
Let $w$ be a trace and $u$ and $v$ two prefixes of $w$.
The set of prefixes common to both $u$ and $v$ has a maximum
(for the prefix relation) called the \emph{longest common prefix} of $u$
and $v$ and denoted $\lcp(u,v)$.
Let $u'',v''$ such that $u=\lcp(u,v)u''$
and $v=\lcp(u,v)v''$.
Then $\dom(u'')\cap \dom(v'')=\emptyset$.
\end{lemma}
\begin{proof}
The proof of the lemma is by induction on $|w|$.

The case where $|w |=0$ is trivial,
in this case $\lcp(u,v)$ is the empty trace.

Assume $|w|\geq 1$.
Denote $L(u,v)$ the set of prefixes common to both $u$ and $v$.
Let $a$ be a letter and $w'$ be a trace such that $w=aw'$.

Assume $a\not \pref u$ and $a\not\pref v$ then $u$ and $v$ are two prefixes of $w'$
and the proof of this case follows by induction.

If $a \pref u$ and $a \not\pref v$ then let $u=au'$.
Then $L(u,v)=L(u',v)$. Since both $v$ and $u'$ are prefixes of
$w'$ then $\lcp(u',v)$ is inductively well-defined.
Since $L(u,v)=L(u',v)$ then $\lcp(u,v)=\lcp(u',v)$ and the proof 
of this case follows by induction.
The case $a \not \pref u$ and $a \pref v$ is symmetric.

If both $a \pref u$ and $a \pref v$ then let $u=au'$ and $v=av'$.
Since both $v'$ and $u'$ are prefixes of
$w'$ then $\lcp(u',v')$ is inductively well-defined. 
Denote $\ell = a \cdot \lcp(u',v')$.
Remark that $a\cdot L(u',v') \subseteq L(u,v)$ thus $\ell \in L(u,v)$
and $\ell$ is a good candidate for $\lcp(u,v)$. 
For that we show that every $z \in L(u,v)$ is a prefix of $\ell$.
There are two cases.
First case: $a\pref z$ thus there exists $z'$ such that $z=az'$ then $az'\pref au'$ and $az'\pref av'$
thus $z'\in L(u',v')$ hence $z'\pref \lcp(u',v')$ thus $z =az' \pref a \cdot \lcp(u',v') = \ell$.
Second case, $a\not \pref z$. Then~\eqref{comm} implies
$\dom(a) \cap \dom(z) = \emptyset$
and
$az \in L(u,v)$.
Thus $z \in L(u',v')$ hence
$z \pref \lcp(u',v')$ and there exists $z' $ such that $\lcp(u',v') = z z'$.
Then $\ell = a \cdot \lcp(u',v') = a z z' = z a z'$
thus  $z\pref \ell$ which terminates the proof of the second case.
Finally, every $z \in L(u,v)$ is a prefix of $\ell$ and $\ell \in L(u,v)$
thus $\lcp(u,v)$ is well-defined.
The second statement follows easily by induction
since $u' = \lcp(u',v') u''$ and $v'= \lcp(u',v') v''$.
\end{proof}

\newcommand{\fakelemma}[2]
{
{\noindent \bf Lemma~\ref{#1}.}{\emph{
#2}
}
}

\section{Properties of locks: proof of Lemma~\ref{lem:lock}}

\fakelemma{lem:lock}{
Let $u$ be a prime play of a game $G$ and $\QQ\subseteq \PP$.
Each of the following conditions is sufficient for $u$ to be a $\QQ$-lock:
\begin{itemize}
\item[i)]
$\QQ=\PP$.
\item[ii)]
$u$  is a $(\PP\setminus \QQ)$-lock.
\item[iii)]
$\QQ\subseteq\dom(\last(u))$.
\item[iv)]
The game is series-parallel and $\QQ=\dom(B)$
where $B$ is the smallest node of the decomposition tree of $A$ which contains
$\alphabet(u)$.
\item[v)]
The game is connectedly communicating game with bound $k$, $\QQ=\dom(u)$ and $\forall p \in dom(u), |u|_p\geq k$.
\item[vi)]
The game is acyclic with respect to a tree $T_\PP$
and $\QQ$ is the set of descendants in $T_\PP$
of the processes in $\dom(\last(u))$.
\item[vii)]
There are two traces $x$ and $z$ such that $u=xz$ and  $z$ is a $\QQ$-lock
in the game $G_x$ identical to $G$ except  the initial state is changed to $\state(x)$.
\end{itemize}
}
\begin{proof}
For the remainder of the proof we fix $v$ a prime play parallel to $u$ and $c=\last(v)$.
We  denote $b=\last(u)$ (thus $c \ind b$ since $v$ is parallel to $u$).
To show that $u$ is a $\QQ$-lock we need to show that $c$ is $\QQ$-safe.
We prove that any of the conditions i) to vii) is sufficient to prove that $c$ is $\QQ$-safe.

\medskip

Condition i) is sufficient because every letter is $\PP$-safe. 

\medskip

Condition ii) is sufficient because an action is $\QQ$-safe iff it is  $\PP\setminus \QQ$-safe.

\medskip

Condition iii) is sufficient because by hypothesis $c \ind b$ thus $(\dom(c) \cap \QQ)\subseteq 
(\dom(c) \cap\dom(b)) =\emptyset$.

\medskip

For series-parallel games (assume iv) holds) we distinguish between two cases.
In case $c\in B$ then $\dom(c) \subseteq \dom(B)=\QQ$
thus $c$ is $\QQ$-safe.
In case $c\not\in B$ then let $C$ be the smallest node of the decomposition tree containing both $B$ and $\{c\}$.
We show that $C$ is a parallel node.
Indeed $C$ contains both $b\in B$ and $c$ thus it is not a singleton hence not a leaf.
By minimality of $C$, one son of $C$ contains $B$ while the other contains $c$.
Then node $C$ cannot be a serial product because
$\dom(b)\cap \dom(c) =\emptyset$ and one son of $C$ contains $b\in B$ while the other contains $c$.
Thus $\dom(B) \cap \dom(c) =\emptyset$ and $c$ is $\QQ$-safe.

\medskip

For connectedly communicating games (assume (v) holds),
we establish that $c$ is $\QQ$-safe 
by showing that $(\dom(c) \cap \QQ \neq \emptyset) \implies (\dom(c) \subseteq \QQ)$.
Let $p \in \dom(c) \cap \QQ$.
Since $u$ and $v$ are parallel there exists a play $w$
such that both $u \pref w$ and $v \pref w$.
Then $\view_p(u)\pref \view_p(w)$ and $\view_p(v) \pref \view_p(w)$.
Since $c=\last(v)$ and $p \in \dom(c)$ then 
$\view_p(v)=v$ thus $v \pref \view_p(w)$, which we reuse later.
Let $w'$ such that $\view_p(w)= \view_p(u)w'$.
By hypothesis, 
$|u|_p\geq k$ thus $|\view_p(u)|_p\geq k$.
By definition of connectedly communicating games,
since $\view_p(w)$ is prime,
$\view_p(w)= \view_p(u)w'$
and $|\view_p(u)|_p\geq k$
 then
$\dom(w') \subseteq \dom(\view_p(u))$ thus  $\dom(\view_p(w)) \subseteq \dom(\view_p(u))$.
Since $v \pref \view_p(w)$, we get $\dom(v) \subseteq\dom(\view_p(u)) \subseteq \dom(u) = \QQ$.
In particular $\dom(c) \subseteq \QQ$ thus $c$ is $\QQ$-safe which terminates the proof in case (v) holds.

\medskip

For acyclic games (assume (vi) holds), we show that $c$ is $\QQ$-safe as follows.
By definition of acyclic games,
the domain of every action is connected in $T_\PP$.
Let $p \in \dom(b)$ be of minimal depth in the tree
$T_\PP$ among the processes in $\dom(b)$.
Then  $\QQ$ is the set of descendants of $p$ in $T_\PP$.
Since $u$ and $v$ are parallel then $c \ind b$
thus $p\not \in \dom(c)$.
Since $\dom(c)$ is connected in $T_\PP$ then either all processes in $\dom(c)$ are descendants of $p$ or none of them are.
In other words $(\dom(c) \subseteq \QQ) \lor (\dom(c) \cap  \QQ =  \emptyset)$ i.e. $c$ is $\QQ$-safe.

\medskip

Now assume property (vii) holds.
We show that there exists $x',x'',z',z'',v'$ such that:
\begin{align}
& x = x'x''\\
& z = z'z''\\
& v = x'z' v'\\
\label{com1}& \dom(v') \cap \dom(x''z'') = \emptyset\\
\label{com2}&\dom(x'') \cap \dom(z') = \emptyset\enspace.
\end{align}
For that let $y$ be the longest common prefix of $u=xz$ and $v$ i.e. $y=\lcp(xz,v)$.
According to Lemma~\ref{lem:lcp},
there exists $y'$ and $v'$ such that
$yy'=xz$, $yv'=v$
and  $\dom(y')\cap \dom(v')= \emptyset$.
Since $yy'=xz$,
according to Lemma~\ref{lem:facto}
 there exists factorizations
$x=x'x''$ and $z=z'z''$ such that
$y=x'z'$ and $y'=x''z''$
and
$\dom(x'') \cap \dom(z') = \emptyset$.

Now we prove that $xz'v'z''$ is a play.
First, $xz=x'x''z'z''=x'z'x''z''$ thus since $xz$ is a play,
$x'z'x''z''$ is also a play.
And by hypothesis $v=x'z'v'$ is also a play.
Set $w=x'z'$ then to summarize both $wx''z''$ and $wv'$ are plays.
Since the processes playing in $x''z''$ and $v'$ are distinct
(cf.~\eqref{com1})
then $wv'x''z''=x'z'v'x''z''$ is also a play.
And according to~\eqref{com1} and~\eqref{com2},
$x'z'v'x''z''=x'z'x''v'z''=x'x''z'v'z''=xz'v'z''$.
Thus $xz'v'z''$ is a play.

Now, we show that $w=z'v'$ and $z=z'z''$
are two parallel prime plays in the game
$G_x$ with initial state $\state(x)$.
Remark first that both $w$ and $z$ are a prefix 
of $wz''=z'v'z''=zv'$.
And since $xz'v'z''=xwz''$ is a play in $G$ (cf supra)
then $wz''$ is a play in $G_x$,
thus both prefixes $w$ and $z$ are plays in $G_x$ as well.
Since $w$ is a suffix of the prime trace $v=x'z'v'=x'w$, it is prime with maximal action $c=\last(v)$.
Since $z$ is a suffix of the prime trace $u=xz$,
it is prime with maximal action $b=\last(u)$.
Hence $w$ and $z$ are two parallel plays in $G_x$.

By hypothesis, $z$ is a $\QQ$-lock in $G_x$
thus $c=\last(w)$ is $\QQ$-safe.
\end{proof}

\section{Taking shortcuts: proof of Lemma~\ref{lem:shortcut}}

\fakelemma{lem:shortcut}{
Let $(x,y)$ be a  useless $\QQ$-repetition in a  strategy $\sigma$.
Let $\Phi:A^*_\equiv \to A^*_\equiv$ and $\tau$ defined by
$\Phi(u)
=
\begin{cases}
&u \text{ if } x \not \pref u\\
&xyu' \text{ if } x \pref u \text{ and } u=xu'\enspace
\end{cases}$
and
\begin{align*}
\forall p \in \PP, \tau_p(u)
=
\sigma_p(\Phi(\view_p(u))).
\end{align*}
Then $\tau$ is a strategy
called the \emph{$(x,y)$-shortcut of $\sigma$}.
And, for every trace $u$,
\be
\label{eq:tauplayapp}
(\text{$u$ is a $\tau$-play})
\iff 
(\text{$\Phi(u)$ is a $\sigma$-play})
\enspace.
\ee
If $\sigma$ is a winning strategy then $\tau$ is winning as well and has a strictly smaller duration.
}
\begin{proof}

That $\tau$ is a strategy follows from the definition: $\tau_p(u)$ only depends on $\view_p(u)$.

Let $b=\last(x)$.
Since $(x,y)$ is a useless $\QQ$-repetition then
\begin{align}
\label{eq:replocks}
&\text{both $x$ and $xy$ are $\QQ$-locks, in particular they are prime,}
\\
\label{eq:repb}
&\last(x)=\last(xy)=b\\
\label{eq:repincl}
&\dom(y)\subseteq \QQ\\
\label{eq:repstates}
&\state(x)=\state(xy)\\
\label{eq:repsigma}
&
\pi\left(\sigma,x,\QQ\setminus\dom(b)\right)
=
\pi\left(\sigma,xy,\QQ\setminus\dom(b)\right)
\enspace.
\end{align}
\begin{proof}[Proof of property~\eqref{eq:tauplayapp}.]
We start with a preliminary lemma.

\begin{lemma}
\label{lem:tutut}
For every $\sigma$-play $xu'$,
\begin{equation}
\label{tutut}
\forall p \in\PP,
\sigma_p(\view_p(\Phi(xu'))
=
\sigma_p(\Phi(\view_p(xu')))\enspace.
\end{equation}
\end{lemma}
\begin{proof}
First notice that:
\begin{align}
\notag x \pref \view_p(xu') &\iff \view_p(xu')=x\view_p(u') \\
\notag &\iff b \pref \view_p(bu')
\\
\label{iffs}&\iff \view_p(xyu')=xy\view_p(u')\\
\notag& \iff xy \pref\view_p(xyu'),
\end{align}
which comes from applications
of~\eqref{eq:primsuff4} and~\eqref{eq:primconcat}
and~\eqref{eq:repb} and property~\eqref{eq:viewdecapp} of views.

To show~\eqref{tutut}, we consider several cases.

\medskip

{\bf First case.
Assume $x\pref \view_p(xu')$.}
Then according to~\eqref{iffs},
$\Phi(\view_p(xu'))=\Phi(x\view_p(u'))=xy\view_p(u') = \view_p(xyu') = \view_p(\Phi(xu'))$
and in this case~\eqref{tutut} holds.

\medskip

{\bf Second case.
Assume $x\not\pref \view_p(xu')$
and
$\dom(\view_p(u'))\subseteq \QQ$.}
Remark first that 
according to~\eqref{iffs},
$b\not \pref \view_p(bu')$ thus according to~\eqref{eq:viewdecapp},
$\dom(b) \cap \dom(\view_p(u'))=\emptyset$
thus 
$\dom(\view_p(u'))\subseteq (\QQ\setminus \dom(b))$.
Since $x\view_p(u')\pref xu'$ and $xy\view_p(u') \pref xyu'$ 
then both $x\view_p(u')$ and $xy\view_p(u')$ are $\sigma$-plays,
hence 
$(\view_p(u'),\sigma(x\view_p(u')))\in \pi\left(\sigma,x,\QQ\setminus\dom(b)\right)$
and
$(\view_p(u'),\sigma(xy\view_p(u')))\in \pi\left(\sigma,xy,\QQ\setminus\dom(b)\right)$
.
Thus according to~\eqref{eq:repsigma},
we get $\sigma_p(x\view_p(u'))=\sigma_p(xy\view_p(u'))$.
Thus~\eqref{tutut} holds since
\begin{align*}
\sigma_p(\view_p(\Phi(xu')))
&=
\sigma_p(\view_p(xyu'))\\
&=
\sigma_p(\view_p(xy\view_p(u')))\\
&=
\sigma_p(xy\view_p(u'))\\
&=
\sigma_p(x\view_p(u'))\\
&=
\sigma_p(\view_p(x\view_p(u')))\\
&=
\sigma_p(\view_p(xu'))\\
&=
\sigma_p(\Phi(\view_p(xu')))
\enspace,
\end{align*}
where the equalities hold because $\sigma$ is a distributed strategy,
according to property~\eqref{eq:viewview} of views
and because $x\not\pref \view_p(xu')$
thus $\Phi(\view_p(xu'))=\view_p(xu')$.

\medskip


{\bf Third case. Assume $x\not\pref \view_p(xu')$
and $\dom(\view_p(u'))\not\subseteq \QQ$.}
We show by contradiction that
$\dom(\view_p(u'))\cap \QQ =\emptyset$.
Otherwise, since $\view_p(u')$ is prime there would exists
some letter $d \in\alphabet(\view_p(u'))$ such that $\dom(d)$
intersects both $\QQ$ and $\PP\setminus \QQ$.
Let $w =\view_d(\view_p(xu'))$.
Remark that $w\neq \epsilon$ and $\view_p(u')\neq \emptyset$.

We show that $w\not \pref x$ by contradiction.
Otherwise $wu' \pref xu'$ hence
$\view_d(\view_p(wu')) \pref \view_d(\view_p(xu'))=w$.
Let $\RR=\dom(\view_p(u'))$.
Since $\view_p(u')\neq \emptyset$ then $p\in \RR$ thus
according to~\eqref{eq:viewdecapp},
$\view_p(wu')=\view_\RR(w)\view_p(u')$.
And since $w$ is $d$-prime and $\dom(d) \subseteq \RR$
then $\view_\RR(w)=w$ thus $\view_p(wu')=w\view_p(u')$.
Then since $w$ is $d$-prime,
$\view_d(\view_p(wu'))=w \view_d(\view_p(u'))$.
But then $\view_d(\view_p(wu')) \pref w$ shown above implies 
$\view_d(\view_p(u'))=\epsilon$, a contradiction with $d \in\alphabet(\view_p(u'))$.

Thus $w \not\pref x$ hence 
$w =\view_d(\view_p(xu'))\pref xu'$ is a prime play parallel to $x\pref xu'$
with maximal action $d$.
However $d$ is not $\QQ$-safe,
contradicting the hypothesis that $x$ is a $\QQ$-lock (cf.~\eqref{eq:replocks}).
Thus $\dom(\view_p(u'))\cap \QQ =\emptyset$.

Since
$\dom(\view_p(u'))\cap \QQ =\emptyset$
then $\RR \cap \QQ=\emptyset$.
Since $\dom(y)\subseteq \QQ$ (cf.~\eqref{eq:repincl})
then $\view_\RR(y)=\epsilon$ thus
$\view_p(xyu')=\view_\RR(xy)\view_p(u')=\view_\RR(x)\view_p(u')=\view_p(xu')$
according to the property~\eqref{eq:viewdecapp}
of views.
Thus~\eqref{tutut} holds since
\begin{align*}
\sigma_p(\view_p(\Phi(xu')))
=
\sigma_p(\view_p(xyu'))
=
\sigma_p(\view_p(xu'))
=
\sigma_p(\Phi(\view_p(xu')))
\enspace,
\end{align*}
where the last equality holds since
$x\not\pref \view_p(xu')$
thus $\Phi(\view_p(xu'))=\view_p(xu')$.
This completes the proof of~\eqref{tutut}.
\end{proof}

\medskip

We prove~\eqref{eq:tauplayapp}
by induction on $u$.
The base case $u=\epsilon$ holds because $\Phi(\epsilon)=\epsilon$
and $\epsilon$ is consistent with every strategy.
Assume~\eqref{eq:tauplayapp} holds for $u$ and all its prefixes and let
$c$ be a letter. We show that~\eqref{eq:tauplayapp} holds for $uc$ as well.

\medskip

We start with the direct implication.
Assume that $uc$ is a $\tau$-play.
We have to show
\begin{equation}
\label{direct}
\text{$\Phi(uc)$ is a $\sigma$-play}.
\end{equation}

Since $uc$ is a $\tau$-play then $u$ is a $\tau$-play thus by induction hypothesis
$\Phi(u)$ is a $\sigma$-play.
And since $uc$ is a $\tau$-play then
\begin{equation}
\label{tutu}
\forall p \in\dom(c), c \in \tau_p(u)\enspace.
\end{equation}

To show~\eqref{direct} we distinguish between three cases.

{\bf First case: assume $x\not \pref uc$.} Then $\Phi(uc)=uc$
then a fortiori $x \not\pref  \view_p(u)$
thus $\Phi(\view_p(u))=\view_p(u)$.
Hence $\forall p \in\dom(c), \tau_p(u)=\sigma_p(\Phi(\view_p(u)))=\sigma_p(\view_p(u))=
\sigma_p(u)$.
Hence according to~\eqref{tutu},
$\forall p \in\dom(c), c \in\sigma_p(u)$.
Since $u=\Phi(u)$ then $u$ is a $\sigma$-play hence by definition
of $\sigma$-plays, $uc$ as well is a $\sigma$-play.
Thus~\eqref{direct} holds in this case.

{\bf Second case: assume $x \pref uc$ and $x \not\pref u$.} Then $x=uc$ and $c$ is the maximal letter of $x$. Then $\Phi(uc)=\Phi(x)=xy$. Since $(x,y)$ is a $\sigma$-repetition then $xy$ is a $\sigma$-play thus~\eqref{direct} holds.

{\bf Third case: assume $x \pref u$.} Let $u'$ such that $u=xu'$.
By induction hypothesis, $\Phi(u)=xyu'$ is a $\sigma$-play thus to show that
$\Phi(uc)=xyu'c$ is a $\sigma$-play, it is enough to prove
\begin{equation}
\label{tutu2}
\forall p \in\dom(c), c \in \sigma_p(\view_p(xyu'))=\sigma_p(\view_p(\Phi(u))\enspace.
\end{equation}
We show first that
\begin{equation}
\label{tutu3}
\forall p \in\dom(c), c\in \sigma_p(\Phi(\view_p(u)))\enspace.
\end{equation}
This holds because $u=xu'$ is a $\tau$-play thus $\view_p(u)\pref u$ as well
is a $\tau$-play
and by induction hypothesis, $\Phi(\view_p(u))$ is a $\sigma$-play.
Thus by definition of $\tau$,
$\tau_p(u) = \sigma_p(\Phi(\view_p(u)))$
thus~\eqref{tutu3} holds according to~\eqref{tutu}.
Then~\eqref{tutu3} and Lemma~\ref{lem:tutut} show that~\eqref{tutu2} holds.  
This completes the proof of the direct implication of~\eqref{eq:tauplayapp}.

\medskip

Now we show the converse implication of~\eqref{eq:tauplayapp}.
Assume that $\Phi(uc)$ is a $\sigma$-play.
We have to show that $uc$ is a $\tau$-play.

There are two cases.
If $x \not \pref u$.
Then $\Phi(u)=u$ thus by induction hypothesis,
$u$ is both a $\tau$-play and a $\sigma$-play.
Moreover $\sigma$ and $\tau$ coincide on $u$ thus 
since $uc$ is a $\sigma$-play then $uc$ is a $\tau$-play as well.

If $x \pref u$. Let $u'$ such that $u=xu'$.
Then both $\Phi(u)=xyu'$ and $\Phi(uc)=xyu'c$ are $\sigma$-plays.
By induction hypothesis,
$xu'$ is a $\tau$-play thus to show that $xu'c$ is a $\tau$-play we shall show
\begin{equation}\label{tutu7}
\forall p \in\dom(c),
c\in\tau_p(xu')=\sigma_p(\Phi(\view_p(xu')))\enspace.
\end{equation}
Since $xyu'c$ is a $\sigma$-play then
\[
\forall p \in\dom(c),
c\in\sigma_p(xyu')=\sigma_p(\view_p(\Phi(xu')))\enspace,
\]
hence~\eqref{tutu7} holds according to Lemma~\ref{lem:tutut}.
This terminates the proof of the converse implication of~\eqref{eq:tauplayapp}.
\end{proof}

\begin{proof}[Proof that $\tau$ is winning.]
Since $\sigma$ is winning, the set of $\sigma$-plays is finite.
According to property~\eqref{eq:tauplayapp} and the definition of $\tau$,
every $\tau$-play is either a $\sigma$-play or is a subword of a $\sigma$-play thus the set of $\tau$-plays is finite as well.
Let $u$ be a maximal $\tau$-play.

If $x\not\pref u$ then $u$ is a maximal $\sigma$-play and since
$\sigma$ is winning $u$ is a winning play.

Otherwise $x\pref u$ and $u$ factorizes as $u=xw$.
Since $(x,y)$ is a useless $\QQ$-repetition
then according to~\eqref{eq:repstates} the global state is the same in $x$ and $xy$.
Since transitions are deterministic,
all processes are in the same state in $xw$ and $xyw$.
According to~\eqref{eq:tauplayapp},
since $xw$ is a maximal $\tau$-play,
$xyw$ is a maximal $\sigma$-play,
and since $\sigma$ is winning,
$\forall p\in\PP,\state_p(xyw)\in F_p$.
Thus
$\forall p\in\PP,\state_p(xw)\in F_p$.
This terminates the proof that  $\tau$ is winning.
\end{proof}

All statements of Lemma~\ref{lem:shortcut} have been proved.
\end{proof}

\section{Decomposability}

\subsection{Connectedly communicating games: proof of Lemma~\ref{lem:pdec}}

\fakelemma{lem:pdec}{Connectedly communicating games
are process decomposable.
}
\begin{proof}
Let $\preceq$ be the inclusion preorder.
We assume $G$ is $k$-connectedly communicating and show that $G$ is $(\preceq,k)$-process decomposable.
Let $xy$ be a prime play of $G$ such that $y$ is $k$-repeating.
We set $\QQ=\dom(y)$ and $z=y$
and show that both conditions in the definition of process decomposability are satisfied.
Let $b=\last(y)$.
Then $\dom(b) \subseteq \QQ$
thus
$(\QQ \setminus \dom(b)) \subsetneq \QQ$
and condition~\eqref{pd} is satisfied.
To show that $\view_b(xy)$ is a $\QQ$-lock in $G$,
notice first that since $xy$ is prime then $xy=\view_{b}(xy)$.

Denote $G_x$ the game identical to $G$ except the initial state is $\state(x)$.
Then according to Lemma~\ref{lem:lock},
$xy$ is a $\QQ$-lock in $G$:
according to v) applied to $G_x$
the play $y$ is a $\QQ$-lock in $G_x$
hence according to vii) of the same lemma,
$xy$ is is a $\QQ$-lock in $G$.
\end{proof}

\subsection{Series-parallel games: proof of Lemma~\ref{lem:adec}}

\fakelemma{lem:adec}
{Series-parallel games
are action decomposable.
}
\begin{proof}
Let $T_A$ be the decomposition tree of $A$.
For every non-empty subset $B\subseteq A$ the set of nodes containing $B$
is a branch of $T_A$.
We denote $B_\uparrow$ the smallest node of this branch and moreover we set $\emptyset_\uparrow=\emptyset$.
The preorder $\preceq$ on $2^A$ is defined as 
$
B \preceq B' \iff B_\uparrow \subseteq B'_\uparrow\enspace.
$

Let $xy$ be a prime play such that $y$ is not empty.
Set $\QQ = \dom(\alphabet(y)_\uparrow)$ and $z=y$.
We show that both conditions in
the definition of action decomposable games are satisfied.
Let $G_x$ be the game obtained by changing the initial state to $\state(x)$.
According to property iv) of Lemma~\ref{lem:lock}, $y$ is a $\QQ$-lock in $G_x$. Hence according to 
property vii) of Lemma~\ref{lem:lock}, $xy$ is a $\QQ$-lock in $G$.
Let
$b=\last(y)$ and
$
A'=  \{ a \in A \mid \dom(a) \subseteq (\QQ \setminus \dom(b)) \},
$
we show that $A'\prec \alphabet(y)$.

We start with a preliminary remark.
Let $C\subseteq A$.
We say that $C$ is connected if 
$C$ induces a connected subset of the dependency graph of the alphabet, i.e. if the graph with nodes $C$ and edges $D \cap C\times C$ is connected.
If $C$ is connected then $C_\uparrow$ is either a product node or a leaf of $T_A$, because if $B$ is the parallel product of $B_1$ and $B_2$ and $C\subseteq B$ then either $C\subseteq B_1$ or $C\subseteq B_2$.

In particular, since $y$ is prime then $\alphabet(y)$ is connected 
thus $\alphabet(y)_\uparrow$ is either the singleton $\{b\}$
 or a serial product node with two sons $B$ and $C$.
In the first case, $\QQ=\dom(b)$ thus $A'=\emptyset \prec \alphabet(y)$.
In the second case w.l.o.g. assume that $b\in B$.
Then no action of $C$ is independent of $b$
thus $A'\subseteq B$.
Then  $B_\uparrow = B \subsetneq \alphabet(y)
 \preceq \alphabet(y)_\uparrow$
thus $B \prec  \alphabet(y)$
hence
$A' \prec  \alphabet(y)$.
\end{proof}

\subsection{A hierarchy: proof of Lemma~\ref{lem:hier}}

\fakelemma{lem:hier}{Every structurally decomposable game is process decomposable
and every process decomposable game is action decomposable.
}
\begin{proof}
For the first implication, 
fix a preorder $\preceq$ on $\PP$ which is monotonic with respect to inclusion
and witnesses the structural decomposability
of the game.
We show that the game is process decomposable
with parameter $(1,\preceq)$.
Let $xy$ be a prime play.
Then the suffix $y$ is prime.
By definition of structural decomposability,
there exists $\QQ \supseteq \dom(y)$
and a letter $b\in\alphabet(y)$ such that
(H1) $(\QQ \setminus \dom(b)) \prec \QQ$
and (H2) $\forall a \in A, (a\ind b \implies a\text{ is $\QQ$-safe})$.
Let $z$ be a prime prefix of $y$ with maximal letter $b$, which exists since $b\in\alphabet(y)$
thus $y$ factorizes as $y=y'by''$ and we can choose $z =\view_b(y'b)$.
Then (H2) implies that $\view_{b}(xz)$ is a  $\QQ$-lock 
and (H1) is exactly condition~\eqref{pd} in the definition of process decomposability
thus all conditions for  process decomposability are met.

Now assume the game is  process decomposable
with parameters $(k,\preceq_\PP)$.
We define the preorder
$\preceq_A$ on $2^A$  by $(B \preceq_A B') \iff (\dom(B)\preceq_\PP \dom(B'))$.
Then every $(k,\preceq_\PP)$ process decomposable game is $(k,\preceq_A)$
action decomposable
because, $\forall b \in A$,
\[
\dom\left(
 \{ a \in A \mid \dom(a) \subseteq \QQ \text{ and $a \ind b$} \}
\right) \subseteq (\QQ \setminus \dom(b))\enspace
\]
and, as a consequence, for every trace $y$,
\begin{multline*}
((\QQ \setminus \dom(b)) \prec_\PP \dom(y))
\implies
( \{ a \in A \mid \dom(a) \subseteq (\QQ \setminus \dom(b) \} \prec_A \alphabet(y))\enspace.
\end{multline*}
\end{proof}

\section{Decidability of decomposability: proof of Lemma~\ref{decdec}}

\fakelemma{decdec}{ Whether a game is decomposable is decidable.
There exists a computable function $\decomp$
from games to integers
such that whenever a game $G$
is $(\preceq,k)$ decomposable
 for some $k$, it is $(\preceq,\decomp(G))$ decomposable.}
\begin{proof}
The definition of locks can be reformulated using the notion of locked states.

\begin{definition}[$\QQ$-locked global states]
A global state $(q_p)_{p\in\PP}\in \Pi_{p\in\PP} Q_p$  is \emph{$\QQ$-locked}
iff for every prime play $v$ starting from
this state
either $\dom(v) \subseteq \QQ$ or $\dom(v) \cap \QQ = \emptyset$.
\end{definition}

We reformulate what it means for a game \emph{not}
to be decomposable, in terms of computations of asynchronous automata.
For every  $b\in A$,
denote $\AA_b$
the automaton identical to $\AA$ except it is restricted to letters 
whose domain do not intersect $\dom(b)$:
other letters are removed from the alphabet and 
the corresponding transitions are deleted.

\begin{lemma}\label{equivlock}
Let $\QQ\subseteq \PP$.
A prime play with last letter $b$ and global state $(q_p)_{p\in\PP}$ is a $\QQ$-lock
iff the global state $(q_p)_{p\in\PP}$
is $\QQ$-locked in $\AA_b$.
\end{lemma}
\begin{proof}
Reformulation of the definition of $\QQ$-locks.
\end{proof}


 \newcommand{\CC}{\mathcal{C}}
 \newcommand{\BB}{\mathcal{B}}
Let $\preceq$ be a preorder on $2^A$ compatible with inclusion.

For every letter $b$ and subset of processes $\QQ$ denote
\begin{align*}
&A_{\QQ, b}
=
 \{ a \in A \mid \dom(a) \subseteq (\QQ \setminus \dom(b) \}
 \enspace\\
&\CC
=
\left\{
\left(B, b, (q_p)_{p\in\PP}\right)
\mid
B\subseteq A, 
b\in B, 
\exists \QQ\supseteq \dom(B),
A_{\QQ, b} \prec B \text{ and $(q_p)_{p\in\PP}$ is $\QQ$-locked in $\AA_b$}
\right
\}\enspace.
\end{align*}

Remark that $\CC$ is computable because
checking whether a global state is $\QQ$-locked
reduces to checking accessibility 
in the graph of the global states of the automaton,
thus
the set of $\QQ$-locked global states
is computable.

Denote $\AA'$ the 
synchronous automaton reading finite words in
$A^*$ 
which computes on-the-fly the  global state of $\AA$
as well as the list $L\subseteq A$ of maximal actions of the current play.
In particular, $\AA'$ can detect whether the current input word is a prime trace, which is equivalent to $|L|=1$.
Denote $Z$ the set of states of $\AA'$
accessible from the initial state by a \emph{prime} trace.
The following properties are equivalent.

 \begin{itemize}
 \item[i)]
 There exists $k\in\NN$ such that
 the game is $(\preceq,k)$ action decomposable.
 \item[ii)]
 There exists $k\in\NN$
 such that for every prime play $xy$,
if $y$ is $k$-repeating
there exists a prime prefix $z\pref y$
such that
$
(\alphabet(y), \last(z), \state(\view_{\last(z)}(xz)))\in \CC\enspace.$
%
\end{itemize}

Property ii) is actually a simple reformulation of i)
 based on Lemma~\ref{equivlock} and the definition of $\CC$.
 We show that it is decidable. For that we characterize it using the notion of \emph{non-decomposability witness}.

Fix some word $x\in A^*$ and $B\subseteq A$. We say that a word $y\in A^*$ is a non-decomposability witness for $(x,B)$
if:
\begin{itemize}
\item[a)]
$\alphabet(y)=B$,
\item[b)]
the trace $(xy)_\equiv$ is prime,
\item[c)]
$y_\equiv$ has no prime prefix $z_\equiv\pref y_\equiv$ such that
\begin{equation}\label{crit}
(\alphabet(y), \last(z_\equiv), \state(\view_{\last(z_\equiv)}(x_\equiv z_\equiv)))\in \CC\enspace.
\end{equation}
\end{itemize}

We show that the language $L_{x,B}$ of non-decomposability witness for $(x,B)$ is a regular language of finite words.
Condition a) is clearly regular.
Condition b) can be checked with $\AA'$,
initialized with $\state(x)$ and the list of maximal actions in $x$:
for $(xy)_\equiv$ to be prime.
there should be a unique maximal action in this modified version of $\AA'$
after reading $y$.
To show that condition c) is regular,
we show that the set of mirror images of words $y$ \emph{not} satisfying c) is regular.
While reading the mirror image of $y$,
the automaton guesses on-the-fly
the sequence of global states and transitions performed by the automaton,
which should end-up in $\state(x)$ once the first letter of $y$ has been read.
The automaton picks non-deterministically at some moment the last letter $c$ of $z_\equiv$ and
simultaneously guesses $q=\state(\view_{c}(x_\equiv z_\equiv))$
under the constraint 
$(\alphabet(y), c, q)\in \CC$.
From then on the automaton keeps reading $y$ backwards
and computes on-the-fly $\view_{c}(x_\equiv z_\equiv)$ 
using the inductive definition of the view, see~\eqref{defview}.
This way the automaton can check that $q$ is equal to
$\state(\view_{c}(x_\equiv z_\equiv))$.

\medskip

Now we show that property ii) is decidable.
Note that $L_{x,B}$ actually does depend only on $B$ (condition a)),
on the set of maximal actions in $x_\equiv$ (condition b)) and on
$(\view_p(x_\equiv))_{p\in\PP}$ (condition c)).
Thus the collection of possible languages $L_{x,B}$ can be explicitely computed
as a finite collection $(L_{x_i,B_i})_{1 \leq i \leq M}$,
together with the corresponding finite collection of automata $(\AA_{x_i,B_i})_{1 \leq i \leq M}$.

And property ii) holds if and only if
for every language $L_{x,B}$,
\[
f(L_{x,B}) = \sup_{y\in L_{x,B}} \min_{b \in B } |y|_b
\]
is finite, in which case we can choose
\[
k = 
1+ \max_{L_{x,B}} f(L_{x,B})\enspace
\]
to satisfy property ii) otherwise for every $k$ we could find a $k$-repeating word $y \in L_{x,B}$,
thus contradicting the definition of 
$k$-decomposability for the prime play $xy$.

Whether $f(L_{x,B})= \infty$
is equivalent to the existence of a computation loop of the automaton  $\AA_{x,B}$ from which a final state is reachable
and which at the same time contains each letter of $B$.
If such a loop exists, there exists one of length at most $|B|$ times the number of states of the automaton.
Thus we can choose $\decomp(G)=1 + |B| \max_{\AA_{x,B}} |\AA_{x,B}|$.

This terminates the proof of lemma~\ref{decdec}.
\end{proof}




\section{New examples}

\subsection{Four players games: proof of Lemma~\ref{lem:fourp}}

\fakelemma{lem:fourp}
{
Four players games are  structurally decomposable.
}
\begin{proof}
Assume $|\PP|=4$,
we show that the game is structurally decomposable for  the pre-order $\preceq$ on $2^\PP$ defined by:
$
\QQ \preceq \QQ' \iff | \QQ | \leq | \QQ' |\enspace,
$
which is monotonic with respect to inclusion.
Let $y$ be a prime trace.
We set
\[
\QQ = \begin{cases}
\dom(y) & \text{ if } |\dom(y)|\leq 2\\
\PP & \text{ if } |\dom(y)|\geq 3
\end{cases}
\]
If $\dom(y)$ contains a single process $\{p\}$ then we set $b=\last(y)$. Then $\QQ=\{p\}$ thus
$\{p\}\setminus \dom(b) = \emptyset \prec \{p\}$
and every letter $a \ind b$ satisfies $p\not\in\dom(a)$ thus is $\{p\}$-safe.

If $|\dom(y)|\geq 2$ 
then, since $y$ is prime, $y$ contains a letter $b$
such that $|\dom(b)|\geq 2$.
Then $\QQ \setminus \dom(b) \prec \dom(y)$:
in case $|\dom(y)|=2$ then $\QQ \setminus \dom(b)=\emptyset$
and otherwise $|\dom(y)|\geq 3$
and
$|\QQ \setminus \dom(b)| \leq 2$.
And every letter $a \ind b$ is $\QQ$-safe,
if $\QQ= \PP$ this is obvious
and otherwise  $\QQ=\dom(b)$
thus $\dom(a) \cap \QQ=\emptyset$.
\end{proof}

\subsection{Five players games: proof of Lemma~\ref{lem:fivep}}

\fakelemma{lem:fivep}{
Let $G$ be a distributed game with five processes.
Assume that the number of actions
that a process can successively play in a row without 
synchronizing simultaneously with two other processes
is bounded.
Then $G$ is process decomposable.
}
\begin{proof}
Let $B$ be the corresponding bound.
Let $\preceq$ the order on $\PP$ which compares cardinality:
$\QQ\preceq \QQ'\iff |\QQ|\leq |\QQ'|$.
Then the game is $(\preceq,B)$ decomposable.
Let $xy$ a prime play such that 
$\forall p\in\dom(y), |y|_p\geq B$.
Then by hypothesis, $y$ has a prime prefix $z$ such that 
$|\dom(\last(z))|\geq 3$.
We set $\QQ=\PP$ then $\view_{\last(z)}(xz)$ is a $\QQ$-lock
and $|\QQ\setminus \dom(\last(z))| \leq 2 < 3 \leq |\dom(y)|$
thus the conditions of process decomposability are satisfied.
\end{proof}

\subsection{Majority games: proof of Lemma~\ref{lem:majority}}

\fakelemma{lem:majority}
{Assume that every non-local action synchronizes a majority of the processes i.e. for every action $a$,
$
|\dom(a)| = 1 \text{ or }   | \dom(a) |  \geq | \PP \setminus \dom(a) |.
$
Then the game is  structurally decomposable.
}
\begin{proof}
We show that the game is structurally decomposable for  the pre-order $\preceq$ on $2^\PP$ defined by:
$
\QQ \preceq \QQ' \iff | \QQ | \leq | \QQ' |\enspace.
$
Let $y$ be a prime trace.

If $\dom(y)$ contains a single process $\{p\}$ then we set $b=\last(y)$ and $\QQ=\{p\}$. Then every letter $a \ind b$ is $\QQ$-safe and $\QQ \setminus \dom(b) = \emptyset \prec \{p\}$.

If $ |\dom(y)| \geq 2$
then, since $y$ is prime, $y$ contains a letter $b$
such that $|\dom(b)|\geq 2$.
By hypothesis
$ | \PP \setminus \dom(b) | \leq | \dom(b) | $.
We set
\[
\QQ=
\begin{cases}
\PP & \text{ if } 2 * |\dom(y)| > |\PP|\\
\dom(y) & \text{ otherwise.}
\end{cases}
\]

In the first case $2 * |\dom(y)| > |\PP|$ then $\QQ=\PP$ and
every action is $\QQ$-safe.
Since  $ | \PP \setminus \dom(b) | \leq | \dom(b) | $ then
$2 * |\dom(b) | \geq |\PP|$ hence
$ 2 (|\PP| - |\dom(b)|) \leq |\PP| < 2 * |\dom(y)|$.
Thus $(\PP \setminus \dom(b))   \prec  \dom(y)$.

In the second case $2 \leq  |\dom(y)|$ and $2 * |\dom(y)| \leq |\PP|$.
Then  $(\QQ \setminus \dom(b)) \prec \dom(y)$ because $\dom(b) \subseteq \QQ =\dom(y)$.
Since $ |\PP| \leq 2 * |\dom(b) | \leq 2 * |\dom(y)| = |\PP|$
then $\dom(b)=\dom(y)=\QQ$  thus every action $a\ind b$
is $\QQ$-safe.
\end{proof}

\subsection{Merging games: proof of Lemma~\ref{lem:merging}}
\fakelemma{lem:merging}
{
Let $G$ be a game,
and $\PP_0, \PP_1\subseteq \PP$
 two set of processes such that $\PP=\PP_0\cup\PP_1$ and for every action $a\in A$,
\[
(\dom(a) \cap \PP_0\neq \emptyset)\land (\dom(a) \cap \PP_1\neq \emptyset)
\implies(\PP_0\cap \PP_1\subseteq \dom(a))\enspace.
\]
If both projections of $G$ on $(\PP_0\setminus \PP_1)$ and $(\PP_1\setminus \PP_0)$ are structurally decomposable then $G$ is structurally decomposable.
}
\begin{proof}
Let $G_0$ and $G_1$ the projections of $G$ on $\PP_0$ and $\PP_1$ and
$\preceq_0,\preceq_1$ some preorders witnessing that $G_0$ and $G_1$ are structurally decomposable.
Let  $\preceq$ be the preorder on $2^\PP$ defined by:
\[
\QQ \preceq \QQ' \iff
\begin{cases}
&\QQ' \cap \PP_0 \neq \emptyset \land \QQ' \cap \PP_1 \neq \emptyset\\
\lor & (\QQ' \subseteq \PP_0\setminus \PP_1) \land (\QQ \cap (\PP_0\setminus \PP_1) \preceq_0 \QQ')\\
\lor & (\QQ' \subseteq \PP_1\setminus \PP_0) \land (\QQ \cap (\PP_1\setminus \PP_0) \preceq_1 \QQ')\enspace.
\end{cases}
\]
which coincides with $\preceq_0$ and $\preceq_1$ on $2^{\PP_0\setminus \PP_{1}}$ and $2^{\PP_1\setminus \PP_{0}}$ respectively
and all sets intersecting both $\PP_0$ and $\PP_1$ are $\preceq$-equivalent and strictly $\prec$-greater than sets in $2^{\PP_0\setminus \PP_{1}} \cup 2^{\PP_1\setminus \PP_{0}}$.
Then $\preceq$ is monotonic with respect to inclusion
because $\preceq_0$ and $\preceq_1$ are.

We show that $G$ is structurally $\preceq$ decomposable.
Let $y$ be a prime trace.

Assume first $(\dom(y)\cap\PP_{0}\neq\emptyset \land \dom(y) \cap  \PP_1\neq\emptyset)$.
We set $\QQ=\PP$.
Since $y$ is prime 
then $y$ has at least one letter $b$ whose domain intersects both $\PP_0$ and $\PP_1$
thus by hypothesis $\PP_{0}\cap \PP_{1}\subseteq \dom(b)$.
Hence by definition of $\preceq$,
$(\PP \setminus \dom(b)) \prec \dom(b) \preceq \dom(y)$.
And every action is $\PP$-safe thus conditions for structural decomposability are fulfilled in this case.

Assume that $\dom(y)\subseteq \PP_0\setminus \PP_1$
(the case $\dom(y)\subseteq \PP_1\setminus \PP_0$ is symmetric).
Since $G_0$ is structurally $\preceq_0$ decomposable,
there exists $\QQ_0\subseteq \PP_0\setminus \PP_1$ and a letter $b$ of $y$ such that:
\begin{align}
\label{eq:la1}
&\forall a \in A, \dom(a) \cap (\PP_0\setminus \PP_1)\neq \emptyset \land a\ind b \implies \text{$a$  is $\QQ_0$-safe in $G_0$}\\
\label{eq:la2}
&\QQ_0\setminus \dom(b) \prec_0 \dom(y)\enspace.
\end{align}
Set $\QQ=\QQ_0 \cup \PP_1$.
 Since $\dom(y)\subseteq (\PP_0\setminus \PP_1)$  
 and $\QQ\cap (\PP_0\setminus \PP_1)=\QQ_0$
 then~\eqref{eq:la2} and the definition of $\prec$
 implies
  $\QQ\setminus \dom(b) \prec \dom(y)$. 
  We show that every letter $a \ind b$  is $\QQ$-safe for that we assume $\dom(a) \cap \QQ \neq\emptyset$ and
  we prove that $\dom(a) \subseteq \QQ$
  or equivalently $\dom(a) \cap (\PP_0\setminus\PP_1) \subseteq \QQ_0$.
  If $\dom(a) \cap (\PP_0\setminus\PP_1)=\emptyset$ there is nothing to prove.
Otherwise since $\dom(a) \cap \QQ \neq\emptyset$
  then $\dom(a) \cap \QQ_0 \neq \emptyset$.
  Moreover
 according to~\eqref{eq:la1}, $a$ is $\QQ_0$-safe in $G_0$ thus since $\dom(a) \cap \QQ_0 \neq \emptyset$
 then 
  $\dom(a) \cap (\PP_0 \setminus \PP_1)\subseteq \QQ_0$ which terminates to prove that every action $a\ind b$ is $\QQ$-safe. Thus $G$ is structurally decomposable.
\end{proof}
\end{document}